\documentclass{aastex}

\shorttitle{SX Phoenicis Stars in NGC 5466} \shortauthors{Jeon et al.}

\begin{document}

\title{SX Phoenicis  Stars in the Globular Cluster NGC 5466 \\
}

\author{Young-Beom Jeon} 
\affil{Korea Astronomy Observatory, Daejeon 305-348, KOREA \\
Email: ybjeon@boao.re.kr}

\author{Myung Gyoon Lee \altaffilmark{1}}
\affil{Astronomy Program, School of Earth and Environmental Sciences,
Seoul National University, Seoul 151-742, KOREA \\
Email: mglee@astrog.snu.ac.kr}

\author{Seung-Lee Kim}
\affil{Korea Astronomy Observatory, Daejeon 305-348, KOREA \\
Email: slkim@kao.re.kr}

\and

\author{ Ho Lee}
\affil{Department of Earth Science Education, Korea National University of
Education, Choongbuk 363-791, KOREA\\
Email: leeho119@boao.re.kr}

\altaffiltext{1}{Visiting Investigator at the Department of Terrestrial Magnetism, Carnegie Institution of
Washington, 5241 Broad Branch Road, N.W., Washington, D. C., USA}

\begin{abstract}

Through time-series CCD photometry of the globular cluster NGC 5466, we have 
detected nine SX Phoenicis stars including three new ones. All the SX Phoenicis
stars are located in the blue straggler region in the color-magnitude diagram
of NGC 5466.
Five of them show clearly double-radial mode features,
 the periods of  which are
well consistent with the theoretical ratio of the first overtone mode
to the fundamental mode (P$_{1H}$/P$_F$).
Normally, it has not been easy to secure a P-L relation of the SX Phoenicis 
stars, because determination of the pulsational mode of the SX Phoenicis stars has been difficult.
The existence of five SX Phoenicis stars in NGC5466 with double-radial modes allows
us to derive reliably a P-L relation for the fundamental mode of  the SX Phoenicis stars.
Using seven SX Phoenicis stars including five stars with double-radial modes,
we derive a P-L relation for the fundamental mode in NGC 5466,
$<V> = - 3.25 (\pm 0.46) Log P + 14.70 (\pm 0.06 ),~ (\sigma = \pm0.04) $,
corresponding to
$<M_V> = - 3.25  (\pm0.46)Log P - 1.30 (\pm0.06 ) $
for an adopted distance modulus of $(m-M)_0=16.00$ and zero reddening. 

\end{abstract}

\keywords{Globular clusters: individual (NGC 5466)) ---
stars: blue straggler stars
          --- stars: oscillations --- stars: variable stars }

\section{Introduction}

NGC 5466 (R.A.\,=\,14$^h$ 05$^m$ 27$\fs$3, Decl.\,=\,+28$\arcdeg$ 32$\arcmin$
04$\arcsec$, J2000.0) is a sparse globular cluster with an extremely low metallicity 
[Fe/H]=--2.22 \citep{har96},
having a large number of blue stragglers. \citet{nem87} found 48 blue straggler stars (BSSs)
which are mostly concentrated within $r=3'$ from the center of the cluster.
Several kinds of variables stars are known to exist in this cluster:
six SX Phoenicis stars \citep{nem90,che99},
three eclipsing binaries \citep{mat90,kal92,mck98},
a Cepheid and more than twenty RR Lyrae stars \citep{cle01}.

In this paper we present a photometric study of SX Phoenicis stars in NGC 5466.
In the long-term search for the short-period variable stars in Galactic globular clusters,
we have found three new SX Phoenicis stars named Cl* NGC 5466 SXP 1, Cl* NGC 5466 SXP 2 and 
Cl* NGC 5466 SXP 3 in addition to the previously known six 
named Cl* NGC 5466 NH 27, Cl* NGC 5466 NH 29 Cl* NGC 5466 NH 35, Cl* NGC 5466 NH 38, 
Cl* NGC 5466 NH 39 and Cl* NGC 5466 NH 49 by \citet{nem90}; 
hereafter referred to as SXP1, SXP2 and SXP3 for new ones, and 
NH27, NH29, NH35, NH38, NH39 and NH49 for known ones, respectively.
Results on the eclipsing binaries, a Cepheid and RR Lyrae stars in NGC5466
will be presented in a separate paper \citep{lee04}.
We adopt the zero interstellar reddening and the distance modulus $(m-M)_0=16.00$ in this study \citep{har96}.

This paper is composed as follows. Observations and data reduction are described in $\S$ 2.
$\S$ 3 describes selection of variable stars in NGC 5466,
and $\S$  4 presents the light curves and frequency analysis of the SX Phoenicis stars.
$\S$  5 discusses the characteristics of these SX Phoenicis stars, including
mode identification and the P-L relation.
Finally primary results are summarized in $\S$ 6.

\section{Observations and Data Reduction}

\subsection{Observations}

We obtained time series CCD images of NGC 5466 for 22 nights from
February 8th, 1999 to March 23rd, 2002.
Total observing time is 27.1 and 133.7 hours for $B$ and $V$ bands,
respectively.
A total of 48 and 944 frames were obtained for $B$ and $V$ bands,
respectively.
Because the observations were performed under various seeing (0.9$\sim$4.5 arcsec)
and weather conditions, we adjusted exposure times depending on the seeing
and transmission of the night sky.
The observation log is listed in Table 1.
\placetable{tbl-1}
The CCD images were obtained with a thinned SITe 2k CCD
camera attached to the  1.8m telescope
at the  Bohyunsan Optical Astronomy Observatory (BOAO) in Korea.
The field of view of a CCD image is $11$\farcm$6 \times 11$\farcm$6$
 ($0.3438$ arcsec pixel$^{-1}$) at the f/8 Cassegrain focus of the telescope.
The readout noise, gain and readout time of
the CCD are 7.0 $e^-$, 1.8e $^-/$ADU and 100 seconds, respectively.

A greyscale map of a $V$ band CCD image is shown in Figure 1.
It shows only
a central region ($7\farcm6\times5\farcm7$) of the cluster, out of the total observing
field of $11\farcm6\times11\farcm6$.
Nine SX Phoenicis stars are represented by circles labeled their names
in Figure 1.
\placefigure{fig1}

\subsection{Data Reduction}

Using the IRAF/CCDRED package, we processed the CCD images to correct
overscan regions, trim unreliable subsections, subtract bias
frames and flatten images. Instrumental magnitudes
were obtained using the point spread function (PSF) fitting photometry
routine in the IRAF/DAOPHOT package \citep{ste87,mas92}.
\citet{nem87} presented photoelectric photometry of 36
stars in the outer region of NGC 5466, of which
we select ten stars for the 
standardization of the instrumental magnitudes.
They are located  in  our observing fields.
For standardization, we selected nine and five frames for $V$ and $B$ bands, respectively,
taken under good seeing conditions 
on March 12, 2002.
The derived transformation equations are
\begin{equation}
   V = v + constant - 0.071(\pm0.01)(B-V),
\end{equation}
\begin{equation}
   B = b + constant + 0.169(\pm0.01)(B-V),
\end{equation}
where $v$ and $b$ are instrumental magnitudes of
$V$ and $B$ bands, respectively.
The color coefficients are average values for nine frames for $V$ band
and five farmes for $B$ band, respectively.
$Constants$ are the zero points of the individual frames.
Finally, we have obtained the standard magnitudes of the stars by  averaging
the magnitudes of all the frames.
The residuals between photoelectric magnitudes and derived
magnitudes for standard stars are $\Delta V = -0.001 \pm 0.027$ and $\Delta (B-V)
= 0.005 \pm 0.056$, respectively. There are no systematic deviations
depending on magnitudes or colors.

\section{Selection of Variable Stars}

In Figure 2 we display ($V$, $B-V$) diagram of a total of about 10,600 stars
in the observing field of NGC 5466.
The left panel shows the color-magnitude diagram (CMD) 
for a central region at  r$ < $1.$'$0,
and the right panel shows the CMD for an outer region at  r$ \ge$1.$'$0.
\placefigure{fig2}
In Figure 2 the main sequence (MS), the red giant branch (RGB)
and the horizontal branch (HB) are clearly seen on both panels.
In addition, there are
about 60 stars at the brighter and bluer region above the MS turnoff
(represented by boxes on both panels), which are blue stragglers.

We applied the ensemble normalization technique
\citep{gil88,jeo01} to normalize instrumental magnitudes between
time-series CCD frames. We used about a hundred  normalizing
stars ranging from 13.7 mag to 19.0 mag for the $V$ band and from 13.7 mag
to 18.5 mag for the $B$ band. We exclude variable stars and
stars located within r = 0.$'$5 to avoid the crowding effects.
For   $B$ band data, we use them only for obtaining mean magnitudes, because
the data quality was not good enough to apply frequency analysis.
The normalization equation we used is
\begin{equation}
  B ~ or ~ V = m + c_1 + c_2(B-V) + c_3P_x + c_4P_y
\end{equation}
where $B$, $V$, and $m$ are the standard  and instrumental
magnitudes of the normalizing stars, respectively. $c_1$ is the zero
point and $c_2$ is the color coefficient. $c_3$ and $c_4$ are
used to correct position-dependent terms such as  atmospheric
differential extinction, flat field error and  variable PSF.
The typical values of the coefficients are  $c_1=-3.265$, $c_2=-0.037$, $c_3=0.000021$ and $c_4=-0.000029$.
$P_x$ and $P_y$ are the positions in the CCD, which are ranged from 1 to 2048.

After photometric reduction  of the time-series frames,
we inspected by eyes luminosity variations for about 10,600 stars in the entire field
to search for variable stars.
From this 
we have detected nine SX Phoenicis stars. Six of them are previously known,
and three  are newly discovered in this research. 
We also recovered previously known
19 RR Lyrae stars, three eclipsing binaries and a Cepheid. 

Among the nine SX Phoenicis stars, 
the three newly discovered ones are designated by SXP1, SXP2 and SXP3.
SXP1 is  located near a bright star, and SXP2 and SXP3 are located very close
to the center of the cluster and have very low amplitudes (see Figure 3).
Although seven of the nine SX Phoenicis stars are located in the central region
at r$<1.'5$, we could detect them easily due to the sparseness of the cluster.
For the globular clusters M53 \citep{jeo03} and M15 \citep{jeo01},
we found SX Phoenicis stars only in the outer region at r$>2.'0$;
those clusters have very crowded central regions, so it is very difficult
to detect SX Phoenicis stars in their central regions.

The coordinates, mean magnitudes and color indices of the nine
SX Phoenicis stars are listed in Table 2.
\placetable{tbl-2}
The right ascension and declination coordinates (J2000.0) of the stars in Table 2
were obtained from the astrometry using the Guide Star Catalogue (Version 1.1).

\section{Light Curves and Frequency Analysis}

Figure 3 displays $V$ band light curves of the SX Phoenicis stars.
\placefigure{fig3}
In each panel we show a mean photometric error of each observing night by an errorbar 
in the left-lower corner.
Although NGC 5466 has a well resolved structure, all the SX Phoenicis stars except for NH39
suffer still from contamination effects due to neighbor stars. Some photometric data
for the four SX Phoenicis stars located near the cluster center
were lost by poor seeing
and/or bad sky conditions caused by moon lights or thin clouds.
The light curves in Figure 3 show  typical characteristics of SX Phoenicis stars, i.e.,
short periods and low amplitudes.

We have performed multiple-frequency analysis to determine pulsating
frequencies of  the nine SX Phoenicis stars, using the discrete Fourier transform
and linear least-square fitting methods \citep{kim95,jeo01}.
Figure 4 displays the power spectra of the light curves for
the nine SX Phoenicis stars.
\placefigure{fig4}
Each panel shows the prewhitening process to derive  each peak in the power spectrum
with window spectra represented in the inner panels. 

Low frequencies detected in all the SX Phoenicis stars in NGC 5466 except for NH49 and SXP2
have resulted
from variable seeing condition and/or drift during long observing runs from 1999 to 2002.
Synthetic light curves obtained from these analyses are
superimposed on the data in Figure 3, and they fit the data well. 
Some unrealistic light curves of NH38 around HJD 2452069.0 
are resulted from high-order fitting to small number of data.

The results of the multiple-frequency analysis for  the nine SX Phoenicis stars
are summarized  in Table 3.
\placetable{tbl-3}
%
The signal-to-noise ratios are defined to be the square root of 
the ratio of the power for each frequency to the average power 
after prewhitening all frequencies.
We assume a real frequency to have the amplitude signal-to-noise ratio larger than 4.0
as done by \citet{bre93}.

During the analysis we detected  many harmonic frequencies and probable nonradial frequencies
in addition to the primary frequencies of each star.
Most of the frequencies 
have been affected by 1 cycle day$^{-1}$ alias effect.
The primary period modes for these SX Phoenicis stars range from 0.0386 days to 0.0552 days,
and the semi-amplitudes of the variability range from 0.023 mag to 0.221 mag.

We divide the sample of SX Phoenicis stars into three groups:
1) Double-radial mode SX Phoenicis stars (SXP2, SXP3, NH35, NH38 and NH39),
2) Single-radial mode SX Phoenicis stars with long-term variations (SXP1 and NH29), and
3) Single-radial mode SX Phoenicis stars without long-term variations (NH27 and NH49).
Individual SX Phoenicis stars in each group are described in appendix.

\subsection{Double-Radial Mode SX Phoenicis Stars}

Double-radial mode stars are very useful to identify the pulsating
modes of SX Phoenicis stars.
SXP3, NH35 and NH39 show the double-radial mode oscillations.
Their period ratios are $P_{1H}/P_F$ = 0.7979, 0.7825 and 0.7826 respectively.
The first radial modes are $f_3$, $f_5$ and $f_3$ of SXP3, NH35 and NH39, respectively.
The double-radial mode features of these three stars are considered to be intrinsic ones.
Interestingly, NH39 shows the combination frequencies of the two radial modes;
$f_4$ and  $f_6$ corresponding to  $f_3 - f_1$ and $f_3 + f_1$, respectively.
For NH35 $f_4$ is a suspected harmonic frequency of $f_5 - f_1$ affected by
 1 cycle day$^{-1}$ aliases.
The period ratios help us to identify their pulsation modes with confidence.
The period ratios of SXP3, NH35 and NH39 are close to the theoretical ratios of the fundamental
and first overtone mode for SX Phoenicis stars with extremely low metal abundance\citep{san01}.
In Figure 5 we compare the period ratios of SXP3, NH35 and NH39 
 with the theoretical period ratios 
for various $Log$ L/L$_\odot$ with Z=0.0001 and M/M$_\odot$=1.0 by \citet{san01}.
\placefigure{fig5}
Figure 5 shows that the two double-radial mode SX Phoenicis stars, NH35 and NH39,
are located on the $Log$ L/L$_\odot$=0.8 line, and that
SXP3 is consistent with the $Log$ L/L$_\odot$=0.6 line.
Using Table 2 of \citet{san01} we estimate the temperatures of all these three stars
to be about 7700K.

SXP2 and NH38 in Table 3 are also suspected double-radial mode pulsators.
But we could not obtain precise frequencies for the secondary radial modes
because of the poor data quality.
Their period ratios $P_{1H}/P_F$ are 0.764 and 0.810, respectively.
These values depart from the theoretical ranges for the ratio of $P_{1H}/P_F$.
The frequencies of the suspected to be secondary radial modes
are probably real ones, according to the amplitude signal-to-noise ratios
(see Table 3) and features of power spectra in Figure 4.
If we 
assume that they are affected by the 1 cycle day$^{-1}$ aliases,
the period ratios of SXP2 and NH38 can be 0.781 and 0.798, respectively, which
are well consistent to the theoretical $P_{1H}/P_F$.
We consider SXP2 and NH38 to be candidates of double-radial mode SX Phoenicis stars.

\subsection{Single-Radial Mode SX Phoenicis Stars with Long-Term Variations }

SXP1 and NH29 show distinct low frequencies,
1.5616 and 0.4268 cycle day$^{-1}$, and their
$V$ amplitudes are 0.076 and 0.158 mag, respectively.
The low frequencies are clearly identified in the power spectra of Figure 4
and in light curves of Figure 3.
Their amplitude signal-to-noise ratios are 8.7 and 18.2, respectively.
We have checked an existence of bad pixels on the CCD images, finding no
bad pixels near these stars.
We propose cautiously that the low frequencies of SXP1 and NH29 are caused by
a nonradial $g$-mode and a contact binary, respectively.
SXP1 is a very interesting pulsator,  if their nonradial mode is real.
This is the first discovery in globular clusters for a pulsator
which possesses  $p$-mode (characterized by an SX Phoenicis star)
and $g$-mode (characterized by a $\gamma$ Doradus star), simultaneously.
As an example for this type of pulsators, 
\citet{han02} found  
the characteristics of $\delta$ Scuti and $\gamma$ Doradus type pulsations
for a field binary HD 209295.
But \citet{hen04} found that a long-term period component of HD 207651, 
1.3598 cycle day$^{-1}$ = 0.73540 day, 
was not resulted from $\gamma$ Doradus type term but the ellipticity effect. 
From the spectoscopic observations an eclipsing period was twice the period
of the long-term period seen in the photometry. 

   Figure 6 is a phase diagram of the long period of 2.3430 days for NH29.
   It shows a distinct light variation. 
   If this is a contact eclipsing binary star, the total period will be about
   two times of the long-term period.
   Otherwise, it could be a g-mode frequency similar to the long-term variation
   of SXP1.
   Unfortunately the data are not good enough for accurate frequency analysis.

\placefigure{fig6}
%
%
%

\section{Discussion}

\subsection{ Characteristics of the SX Phoenicis Stars }

In Figure 7, we show the position of the nine
SX Phoenicis stars in the color-magnitude diagram of NGC 5466, listing
their mean magnitudes and color indices in Table 2.
Figure 7 shows that all the  SX Phoenicis stars are located
in the blue straggler region, brighter and bluer than the main sequence turnoff point.
It is known that all the known SX Phoenicis stars in globular clusters are BSSs,
but there are non-variable stars among the BSSs.
%
\placefigure{fig7}

In Figure 8 we have compared the $V$ amplitudes and periods of
the SX Phoenicis stars in NGC 5466 (filled circles) with
those of SX Phoenicis stars in other globular clusters, field SX Phoenicis stars
and $\delta$ Scuti stars.
\placefigure{fig8}
It includes  the SX Phoenicis stars in M53 (star symbols) and M15 (a cross)
discovered by our previous searches for variable blue
stragglers in globular clusters \citep{jeo01,jeo03}.
The sources of the data in Figure 8
are \citet{rol20}  for field SX Phoenicis stars and  $\delta$ Scuti
stars, and \citet{rod20}  for SX Phoenicis stars in  Galactic
globular clusters.
Figure 8 shows that the $V$ amplitudes and
periods of  the SX Phoenicis stars in NGC 5466
are consistent with those for SX Phoenicis
stars in  other globular clusters; the $V$ amplitudes increases steeply with
increasing period, 
and the $V$ amplitudes of SX Phoenicis stars are
larger than those of  $\delta$ Scuti stars with
the same period.

\subsection{Radial Mode Identification}

A method for mode identification  is to use a pulsation constant, $Q$,
denoted as $Q = P \rho^{1/2}$ where $P$ and $\rho$ are the period and mean density
of variable stars, respectively.
$Q$ values of overtone mode are smaller than that of the fundamental mode \citep{bre00}.
$Q$ values can be derived 
using a photometric method such as Str\"{o}mgrem {\it uvby H$_{\beta}$} 
observation \citep{rod03}.
Another photometric method is to use SX Phoenicis stars with double-radial mode,
but the SX Phoenicis stars with double-radial mode are rare.
Fortunately, five (including two candidates) of the SX Phoenicis stars in NGC5466
are found to have double-radial modes, so their modes are identified clearly.
All the double-radial mode stars (SXP2, SXP3, NH35, NH38 and NH39)
show the fundamental and first overtone modes,
as described in $\S$ 3.1.

The mode of the rest of the SX Phoenicis stars can be identified using the amplitude,
period and luminosity.
NH27 is identified as the first overtone mode
because of  its $V$ amplitude, 0.246 mag, being slightly lower,
and especially the position in the period-magnitude diagram as shown in Figure 9.
SXP1 has a high amplitude, and is located in the fundamental region in the
period-magnitude diagram. So it 
is a fundamental mode star.
In Table 3, NH35, NH38, NH39 and NH49 show harmonic frequencies supporting
asymmetric sinusoidal features. Moreover their total light amplitudes have a range of
0.282 $\sim$ 0.730 mag, which are characteristics of high amplitude $\delta$ Scuti (HADS) stars.
So, their mode are clearly identified to be fundamental radial oscillations.
Their primary and secondary radial periods correspond to the fundamental
and first overtone radial modes, respectively.
NH29 
is the first overtone mode star according to its low amplitude (0.142 mag),
well defined sinusoidal feature in the light curve (see Figure 3)
and the location in the period-magnitude diagram.
Seven of nine SX Phoenicis stars in NGC 5466 show
the fundamental mode or double-radial mode with the fundamental and first overtone modes.
They  are displayed by open and filled circles in the color-magnitude diagram in Figure 7, 
respectively.
Two filled triangles denote the first overtone mode stars. 
The identified radial modes are listed in the last column of Table 3.

\citet{mcn20,mcn01} 
suggested that the light amplitude and the degree of asymmetry
of the light curves 
are useful parameters for identifying the pulsating modes.
Generally this method is very useful for systems with a small number of pulsators
and/or no double-radial mode stars therein.
However, in many cases these criteria do not give a unique solution
to identify pulsating modes, especially for SX Phoenicis stars with
very low amplitudes and short periods such as SXP2 and SXP3 in NGC5466.

\subsection{ Period-Luminosity Relation }

The P-L relation of SX Phoenicis stars in the globular
clusters is very useful to obtain the distance moduli of the clusters and nearby
galaxies \citep{mcn95}.
However, it is not easy to define well the P-L relation from the observations,
because there are often a mixture of different pulsation modes \citep{jeo03,mcn01}.
But we can easily identify the pulsating mode using the double-radial mode stars
as described in the previous section.

Figure 9 presents the P-L relation for the SX Phoenicis stars in NGC 5466.
 \placefigure{fig9}
It 
shows that the sample can be
separated into two discrete groups;
one is identified as a fundamental mode group (filled circles),
and the other is a first overtone 
mode group (filled triangles and open triangles).
The solid line represents the fundamental P-L relation,
and the line is shifted to the dashed line
corresponding to the first overtone mode stars by the ratio $P_{1H}/P_F = 0.783 $.
The P-L relation for fundamental mode in NGC 5466 is derived to be
\begin{equation}
  <V> = - 3.25 (\pm 0.46) Log P + 14.70 (\pm 0.06 ),~ (\sigma = \pm0.04),
\end{equation}
which corresponds to
\begin{equation}
<M_V> = - 3.25  (\pm0.46)Log P - 1.30 (\pm0.06 )
\end{equation}
for an adopted distance modulus of $(m-M)_V=16.00$ and $E(B-V)=0.00$ \citep{har96},

The empirical P-L relations have been obtained using field HADS stars and/or
cluster SX Phoenicis stars identified by their pulsating modes.
The slopes derived by field HADS stars and SX Phoenicis stars in
$\omega$ Cen show on the whole steeper ones up to --4.66
\citep{mcn20},
whereas $\delta$ Scuti stars, SX Phoenicis
stars in other globular clusters and theoretical results show flatter slopes up to --2.88
\citep{pyc01}.

The slope of $-3.25$ for NGC 5466 is in a good agreement
with the results for M53 \citep{jeo03} and M55  \citep{pyc01}.
\citet{jeo03} obtained a slope of
$-3.01$ for the fundamental mode stars in M53 and
\citet{pyc01} derived a slope of
$-2.88$ for the fundamental mode and a slope of $-3.1$ for the first overtone mode stars in M55.
The slope for NGC 5466 derived in this study
agrees also well with the theoretical values
of --3.04 by \citet{san01} and --3.05 by \citet{tem02}.

\section{Summary}

Through time-series CCD photometry of the globular cluster NGC 5466,
we detect three newly discovered and six known SX Phoenicis stars.
Owing to the extremely open, well-resolved structure of NGC 5466,
we could detect easily  many short period
variable stars such as SX Phoenicis stars and eclipsing binaries even in the central region.
All the SX Phoenicis stars are found to be located in the blue straggler star
region of the CMD.
Physical parameters of these stars are summarized in Tables 2 and 3.
From the Fourier analysis, we find five double-radial mode stars including two candidates.
These stars are very useful to determine the radial modes of SX Phoenicis stars
in NGC 5466.
These stars 
show  period ratios of the two radial modes 
which are consistent with the theoretical ratio of the first overtone mode
to the fundamental mode (P$_{1H}$/P$_F$).
Using seven SX Phoenicis stars which are considered to be pulsating in the
fundamental mode, 
we derive a P-L relation for the fundamental mode in NGC 5466,
$<V> = - 3.25 (\pm 0.46) Log P + 14.70 (\pm 0.06 ),~ (\sigma = 0.04) $.
The slope of $-3.25$ for NGC 5466 is in a good agreement with the
empirical results for M53 ($-3.01$; \citet{jeo03})
and M55  ($-2.88$; \citet{pyc01}), and the theoretical results
of --3.04 by \citet{san01} and --3.05 by \citet{tem02}.


\acknowledgments
M.G.L. was supported in part by the Korean Research Foundation Grant
(KRF-2000-DP0450).

\appendix
{\bf SXP1:}
   It is located very close to a bright star.
   We could detect four frequencies including a low frequency $f_2=1.5616$
   cycle day$^{-1}$.
   $f_3$ is clearly a harmonic frequency of $f_1$ corresponding to 2$f_1$.
   A close frequency $f_4$ for $f_1$ seems to be a nonradial frequency affected by
   1 cycle day$^{-1}$ aliases.

{\bf SXP2:}
   During  the pre-whitening processes, we have detected five more frequencies whose
   amplitude signal-to-noise ratio is larger than 3.6.
   But we could only confirm three frequencies with their amplitude signal-to-noise ratios larger
   than 4.0 (see Table 3).
   The former two frequencies, $f_1$ and $f_2$ seem to be a pair of double-radial mode,
   and the latter $f_3$ is a combination frequency corresponding to  $f_1 + f_2$.
   Probably $f_3$ was affected by 1 cycle day$^{-1}$ aliases.
   The superimposed synthetic light curves of SXP2 in Figure 3
   are calculated using $f_1$ and $f_2$ only.
   In Figure 4, some frequency peaks are shown in the lower panel of SXP2.
   To confirm these frequencies, better data are needed.

{\bf SXP3:} 
   During  the pre-whitening processes we have detected six frequencies whose
   amplitude signal-to-noise ratios are larger than 3.7.
   But we can only confirm three frequencies:
    $f_1 = 25.8994 $, $f_2=25.4323$ and $f_3 = 32.7041 $ cycles day$^{-1}$.
   Two frequencies, $f_1$ and $f_3$, might be a pair of double-radial mode,
   and $f_2$ be a nonradial mode affected by 1 cycle day$^{-1}$ alias effect.
   Among the rest of six frequencies a peak of 28.1463 cycles day$^{-1}$ in Figure 4
   seems to be an intrinsic one. But the amplitude signal-to-noise ratio is only 3.7.
   To confirm this frequency, better data are needed.

{\bf NH27:}
   NH27 is the brightest star among the nine SX Phoenicis stars in NGC 5466.
   We have detected four  frequencies with their amplitude signal-to-noise ratios larger than 4.0
   (see Table 3).
   $f_2$ and $f_3$ show close frequencies to $f_1$.
   The primary frequency $f_1$ has a higher priority of
   a radial mode frequency among three close frequencies considering the higher amplitude
   and the existence of a harmonic frequency. 
   But the primary frequency $f_1$ is also latent nonradial mode with the other two
   close frequencies.
   Though $f_2$ is a lower frequency (i.e. a longer period) than that of the primary frequency
   $f_1$, it seems to be a nonradial mode.
   $f_3$  can be explained by the excitation of a nonradial mode, too.
   These closely separated nonradial mode frequencies were found in several recent  observations
   of the SX Phoenicis stars \citep{jeo01,jeo03}. 
   The amplitude signal-to-noise ratio of $f_4$ is only 4.0, but its frequency 39.4356
   cycles day$^{-1}$  is identical with 2$f_1$. 
   In Figure 4 the fourth frequency of NH27 is shown to be distinct. 
   It 
   is probably a harmonic frequency of the primary frequency $f_1$.
   We can see easily the asymmetric shape of the harmonic frequency
   and amplitude modulating feature by closely separated frequencies
   in Figure 3.  The frequency ratios are $f_2$/$f_1$ = 0.985 and $f_1$/$f_3$=0.980.

{\bf NH29:}
   The oscillating feature of NH29 is very abnormal.
   It shows a distinct long period variable feature as shown in Figure 3.
   The period of long-term variation is 2.3430 days (see Figure 6).
   $f_2$ of NH29 is a radial frequency with total $V$ amplitude of 0.138 mag.

{\bf NH35:}
   Five frequencies with the amplitude signal-to-noise ratio $\ge$ 5.0
   are detected from multiple-frequency analysis for NH35.
   $f_1$ and $f_5$ are radial mode ones described in $\S$ 3.1.
   $f_2$,  $f_3$ and $f_4$ are harmonic or combination frequencies corresponding to
   2$f_1$, 3$f_1$ and $f_5-f_1$, respectively.
   The total $V$ amplitude is 0.730 mag. 
   This star shows a typical feature of HADS stars
   with only one or two stable frequencies and high amplitude (typically
   $\Delta V \ge 0.4$ mag) compared to the feature of
   complicated oscillation pattern and several frequencies with low amplitude
   (typically $\Delta V \le 0.05$ mag) of
   low amplitude $\delta$ Scuti (LADS; \citet{pet96}) stars.

{\bf NH38:}
   NH38 is located near a brighter star as seen in Figure 1,
   so observing data obtained with bad seeing conditions were not useful.
   Four frequencies are detected for NH38.
   The primary and third frequencies, $f_1$ and $f_3$,
   show a possibility of double-radial mode pair explained $\S$ 3.1.
   $f_2$ and $f_4$ seem to be harmonic frequencies of $f_1$ corresponding to
   2$f_1$ and 3$f_1$.
   In Figure 4 there are two more frequencies peaked on the left side of $f_3$, but their amplitude
   signal-to-noise ratios are too low to confirm them.

{\bf NH39:}
    NH39 is isolated well, so hardly affected by seeing conditions.
    We detect six frequencies with the amplitude signal-to-noise ratio $\ge$ 4.1.
    $f_1$ and $f_3$ are identified radial mode frequencies by the their period ratio.
    $f_2$,  $f_4$ and $f_6$ are  harmonic or combination frequencies corresponding to
    2$f_1$, $f_3 - f_1$ and $f_3+f_1$, respectively.
    A closely separated frequency of $f_1$ is detected by $f_5$ with the amplitude
    signal-to-noise ratio of 5.1.
    NH39 also shows a characteristics of  HADS stars, like NH35,
    with the total $V$ amplitude, 0.460 mag.

{\bf NH49:}
   Comparing to $<$$V$$>$ magnitudes, period and $V$ amplitudes of the primary periods \citep{nem90},
   this 
   is identified as NH49.
   NH49 has a radial mode frequency $f_1$ and a harmonic frequency $f_2$.
   This star 
   is a mono-periodic pulsator.


\clearpage

\plotone{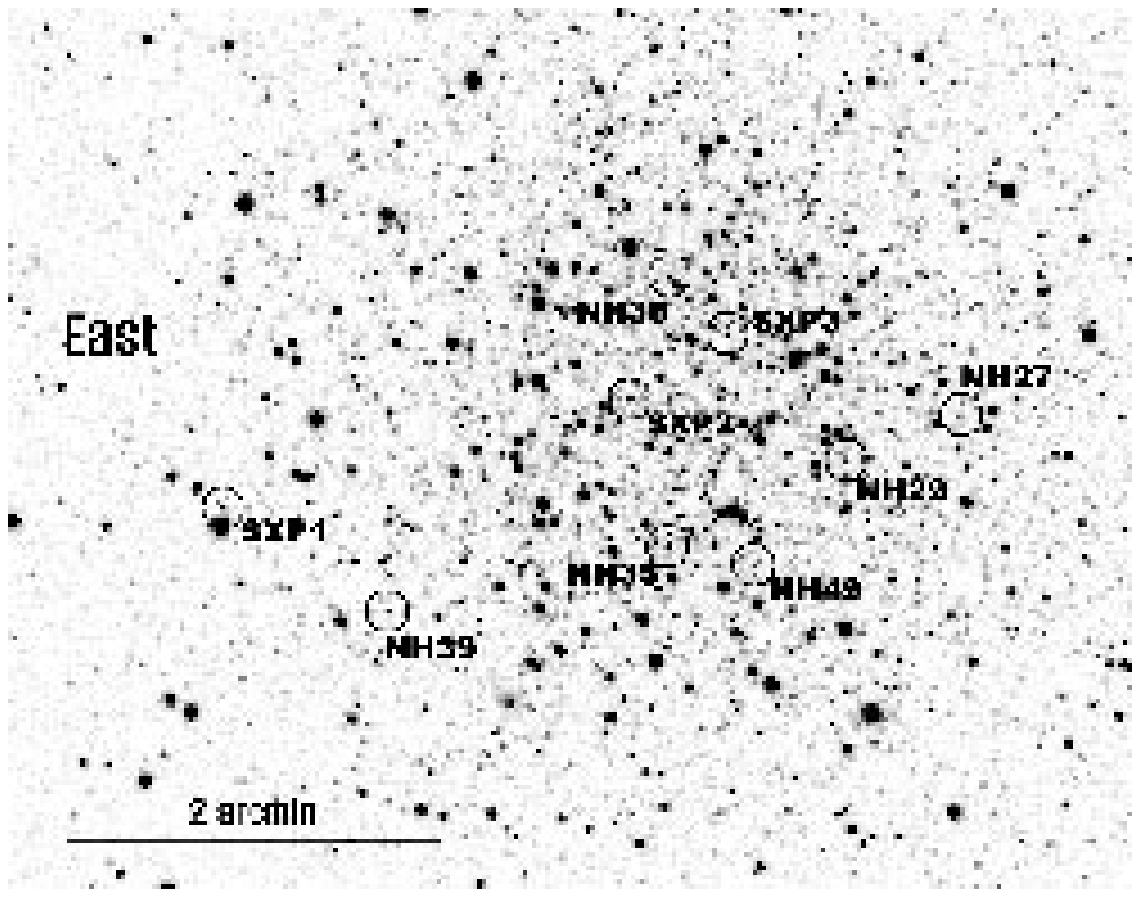}
\figcaption[Jeon.fig01.ps]{ A greyscale map of a $V$-band CCD image of the globular cluster NGC 5466.
The image is presented only for the
central region ($7\farcm6\times5\farcm7$) of the cluster, out of the total observing field of
$11\farcm6\times11\farcm6$.  Nine  SX Phoenicis stars are labeled their names.
\label{fig1}}

\plotone{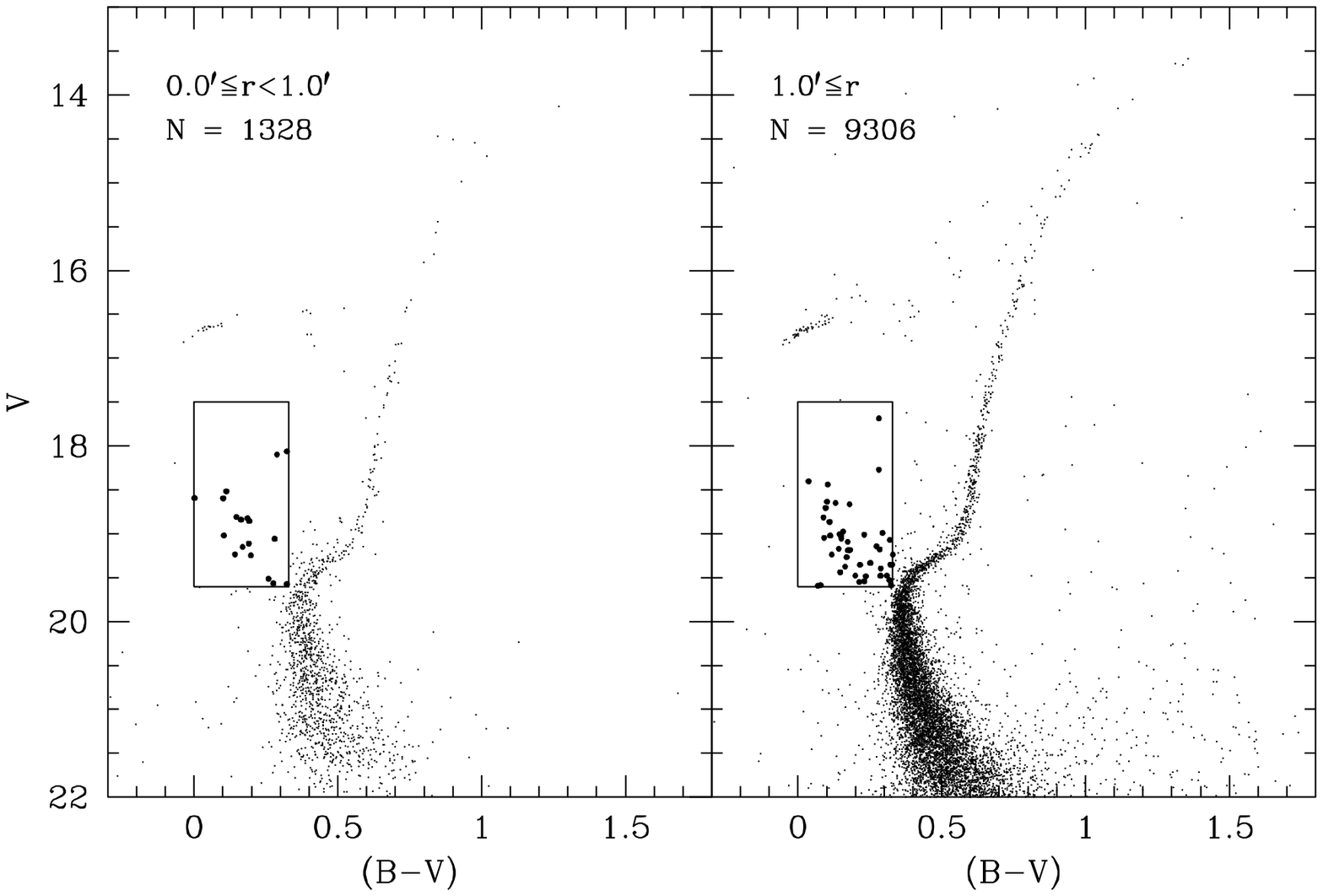}
\figcaption[Jeon.fig02.ps]{
Color-magnitude diagrams of NGC5466.
The left panel is for a central region at r$ < $1$\farcm0$
and the right panel is for an outer region at r$ \ge$1$\farcm$0.
A blue straggler region is outlined by a box.
\label{fig2}}

\plotone{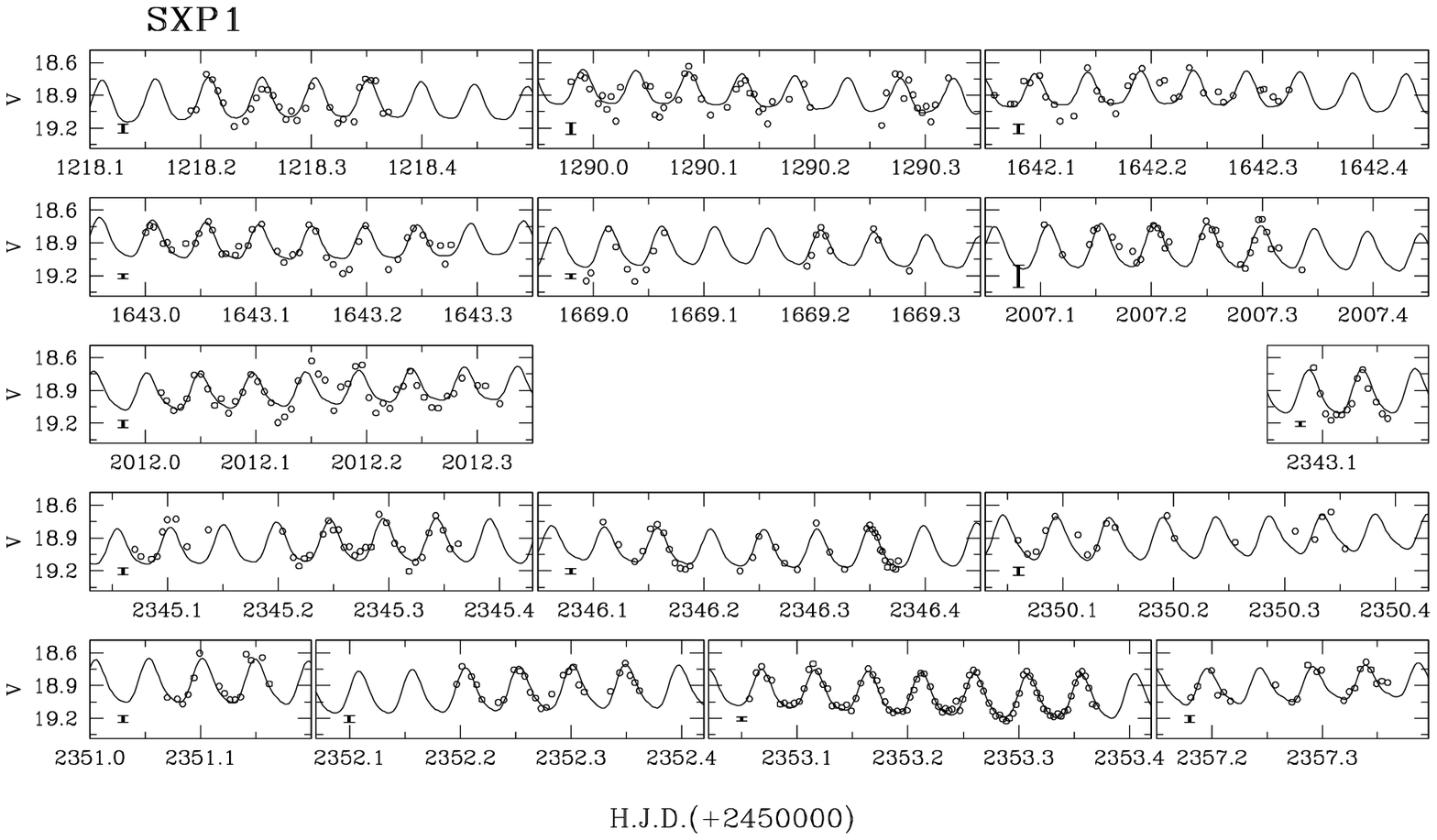}
\plotone{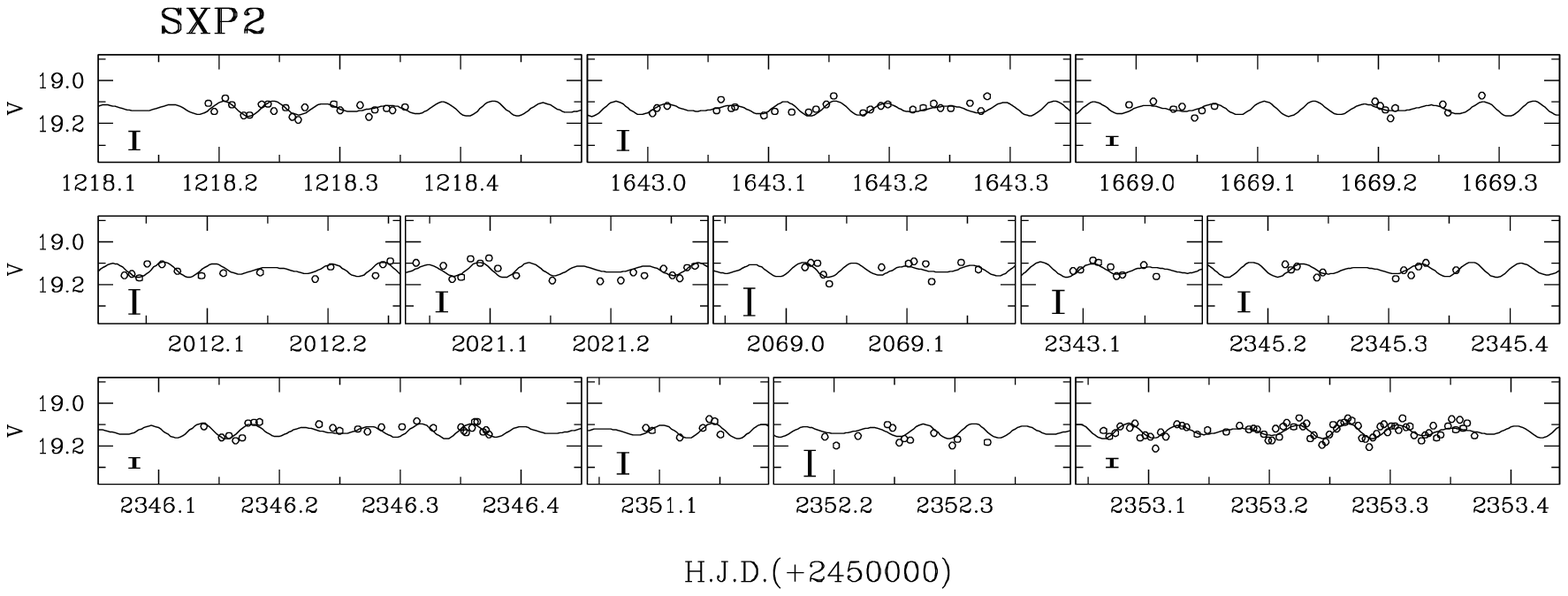}
\clearpage
\plotone{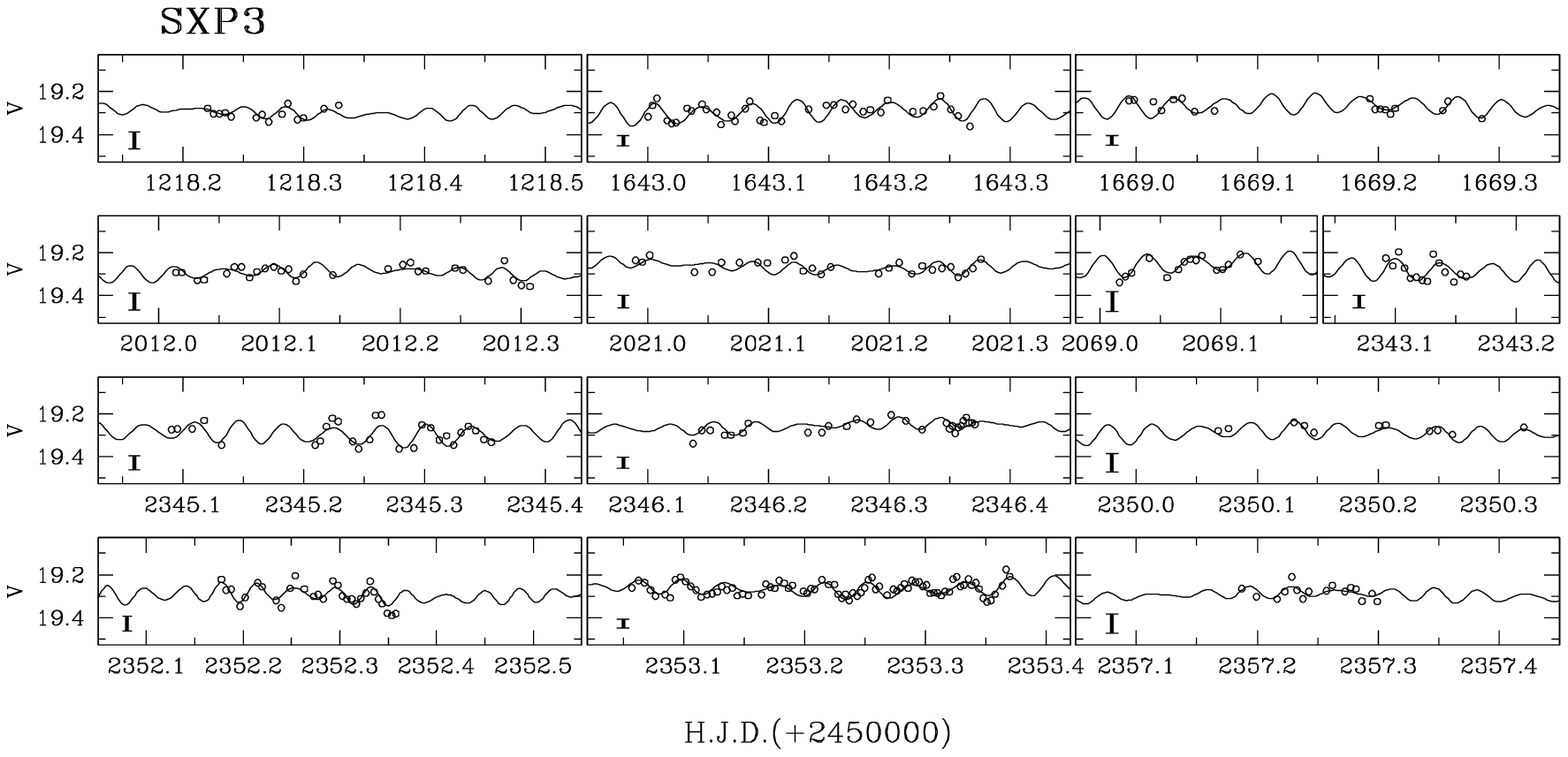}
\plotone{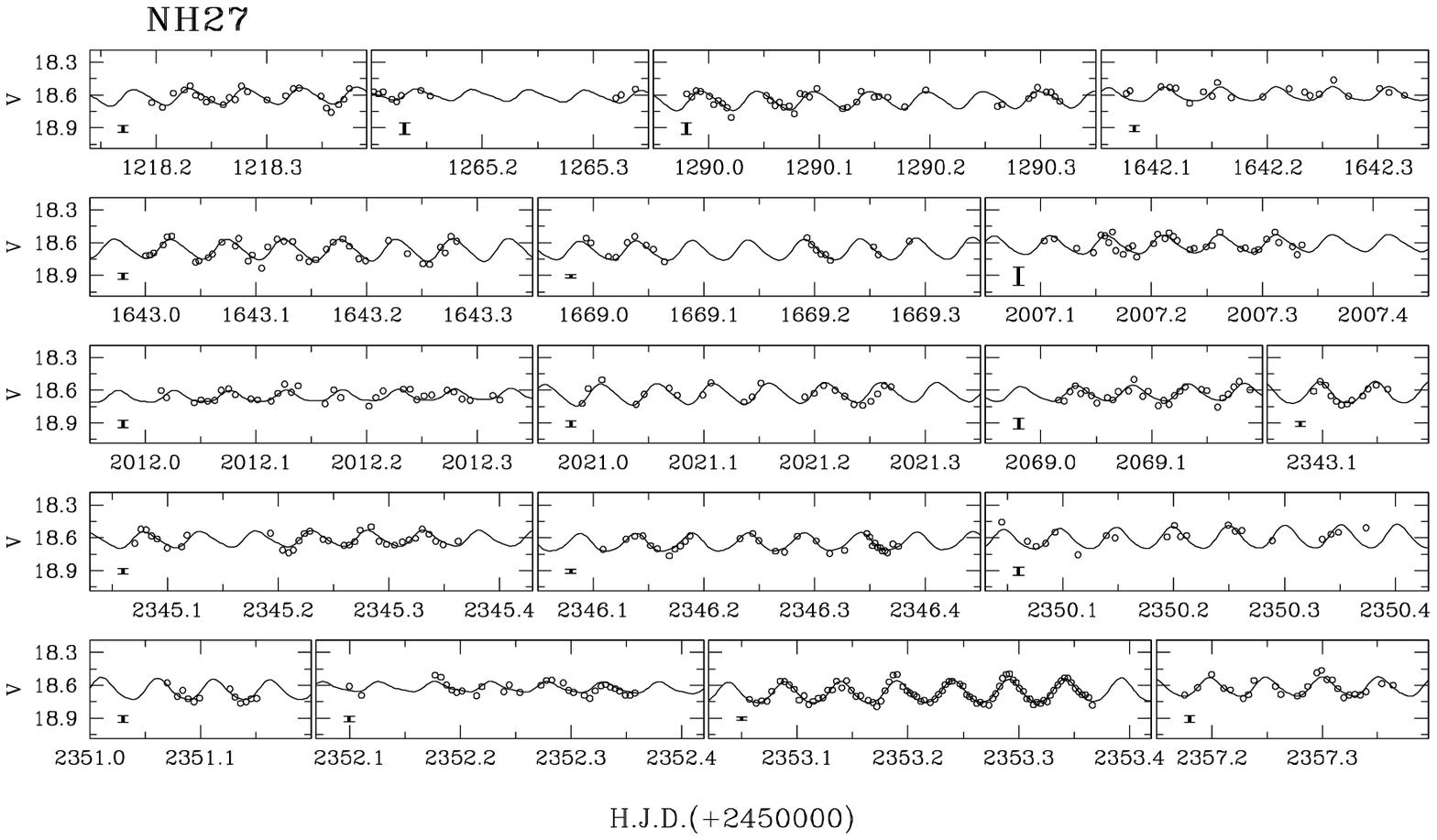}
\clearpage
\plotone{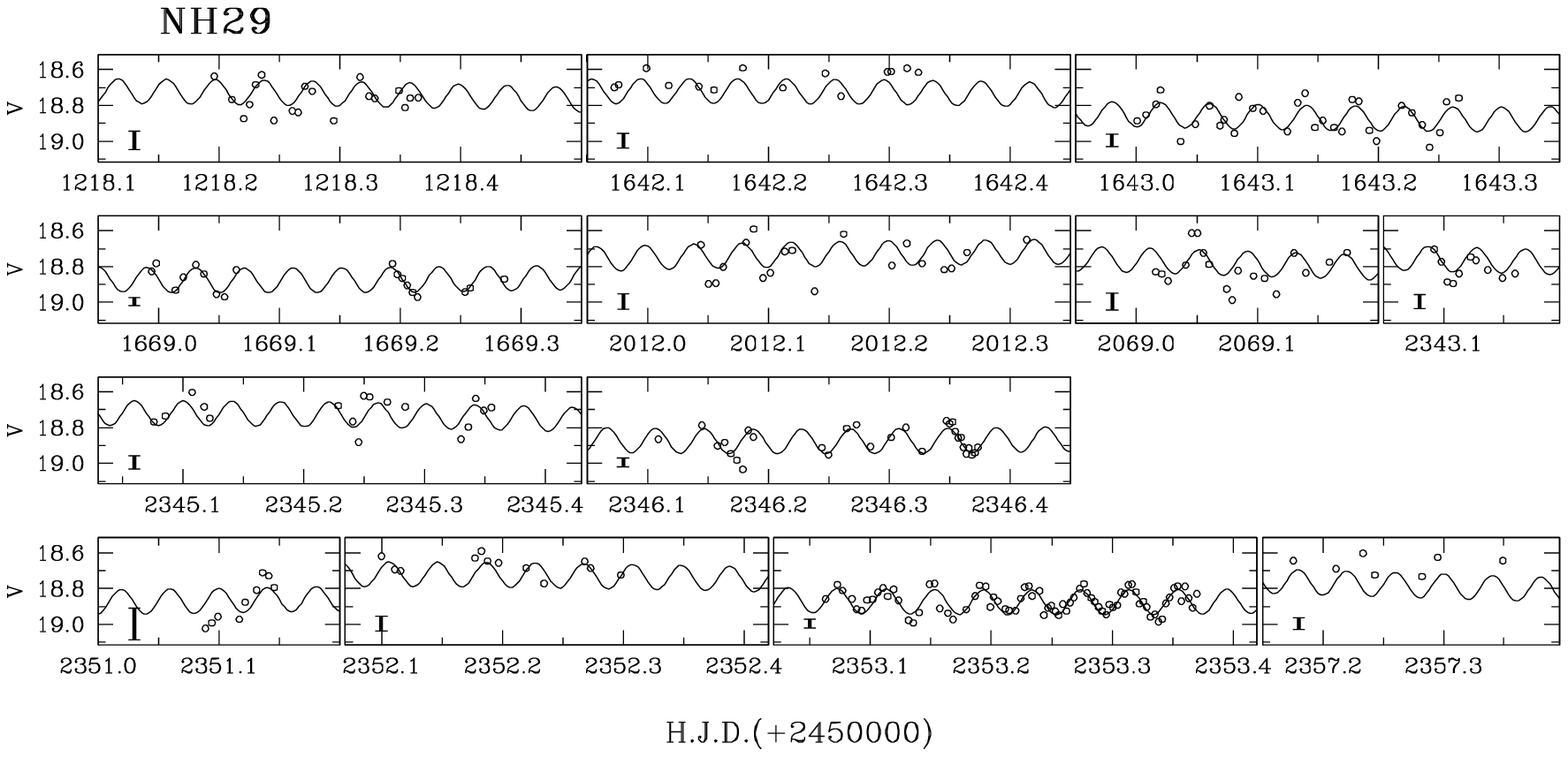}
\plotone{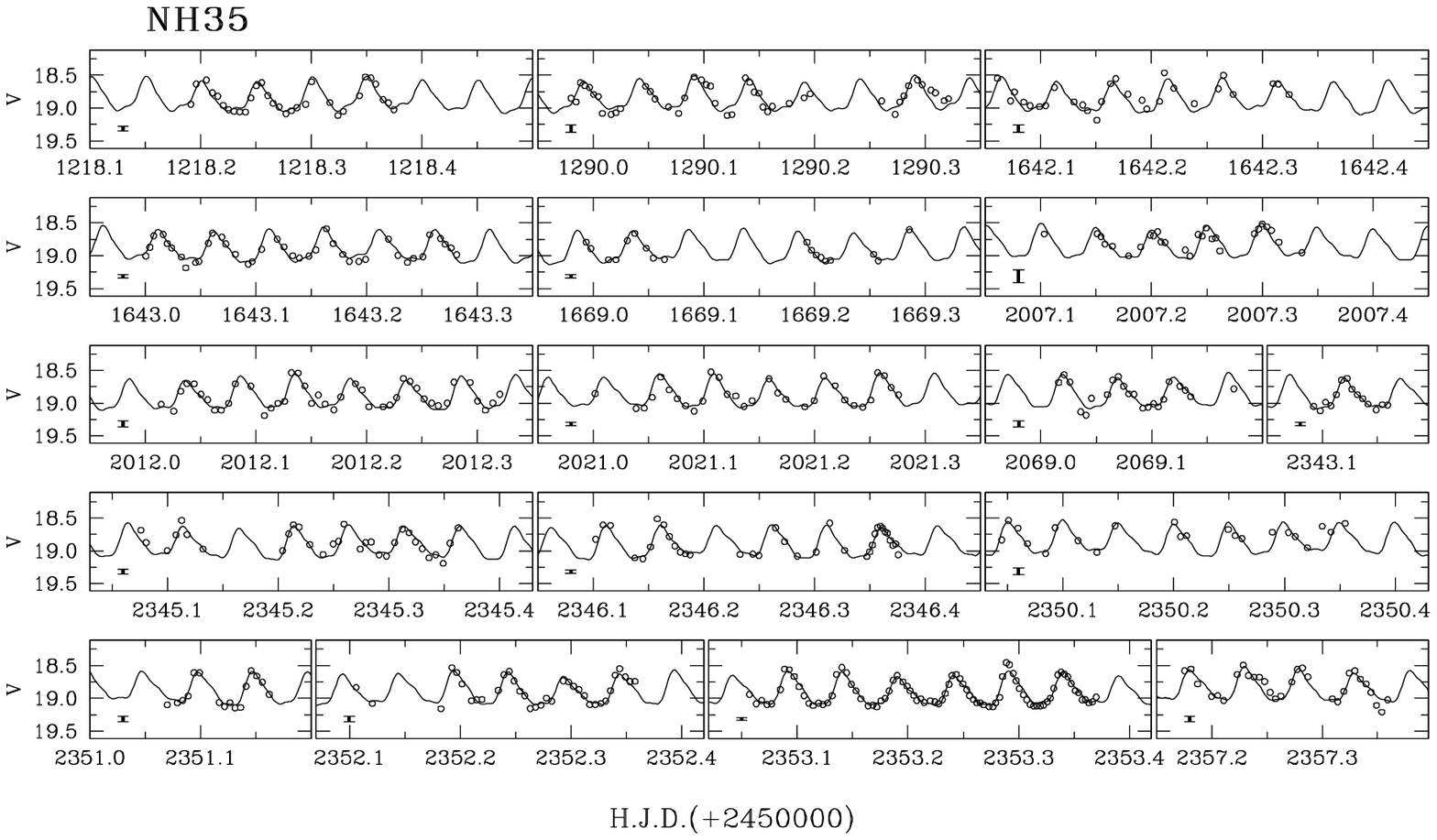}
\clearpage
\plotone{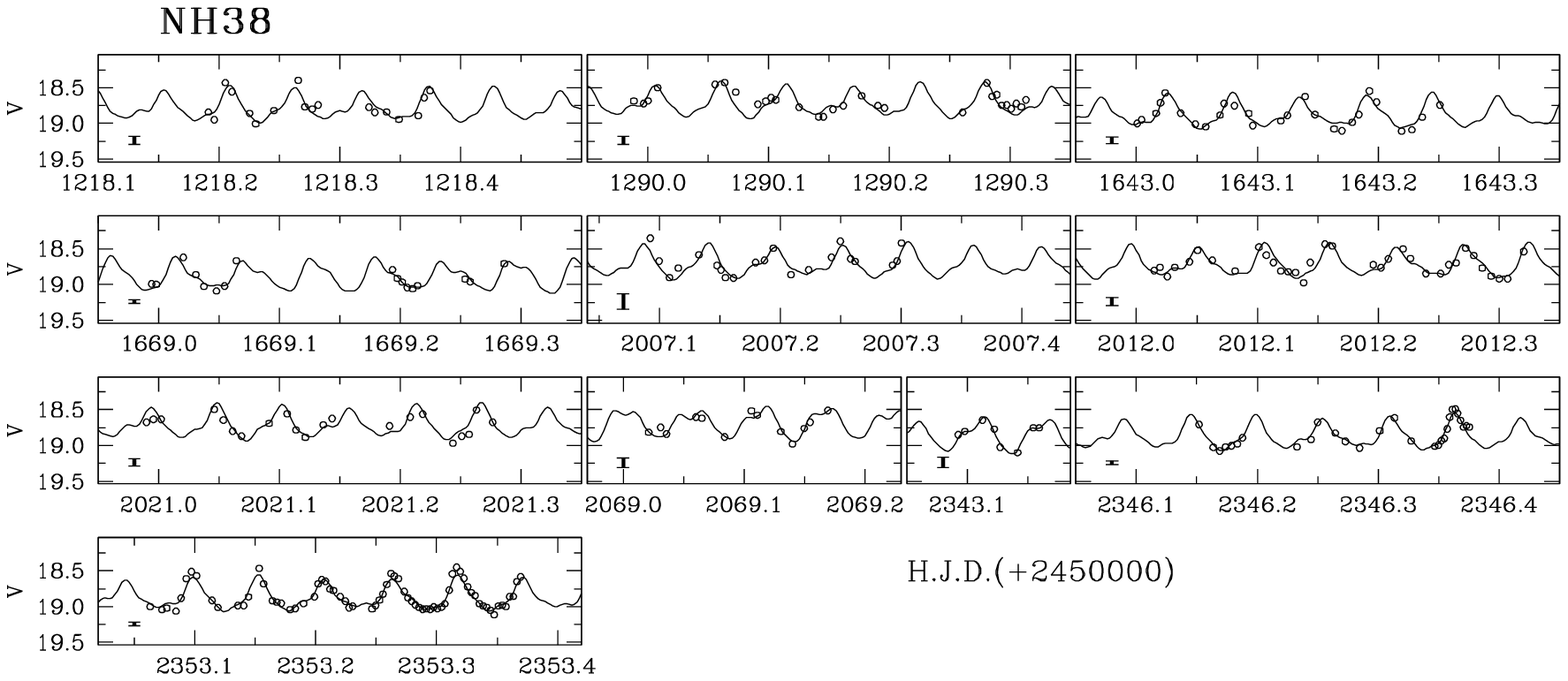}
\plotone{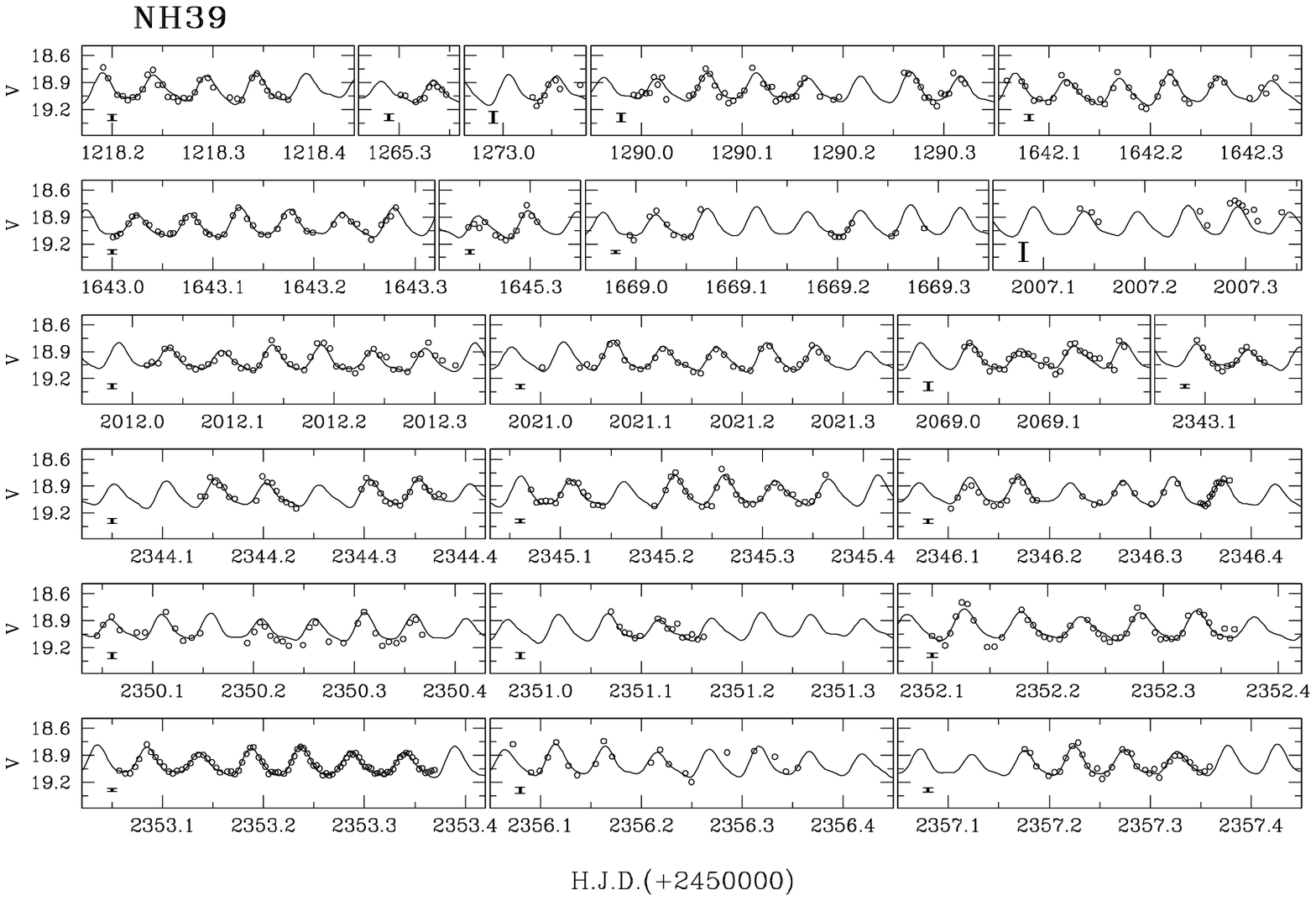}
\plotone{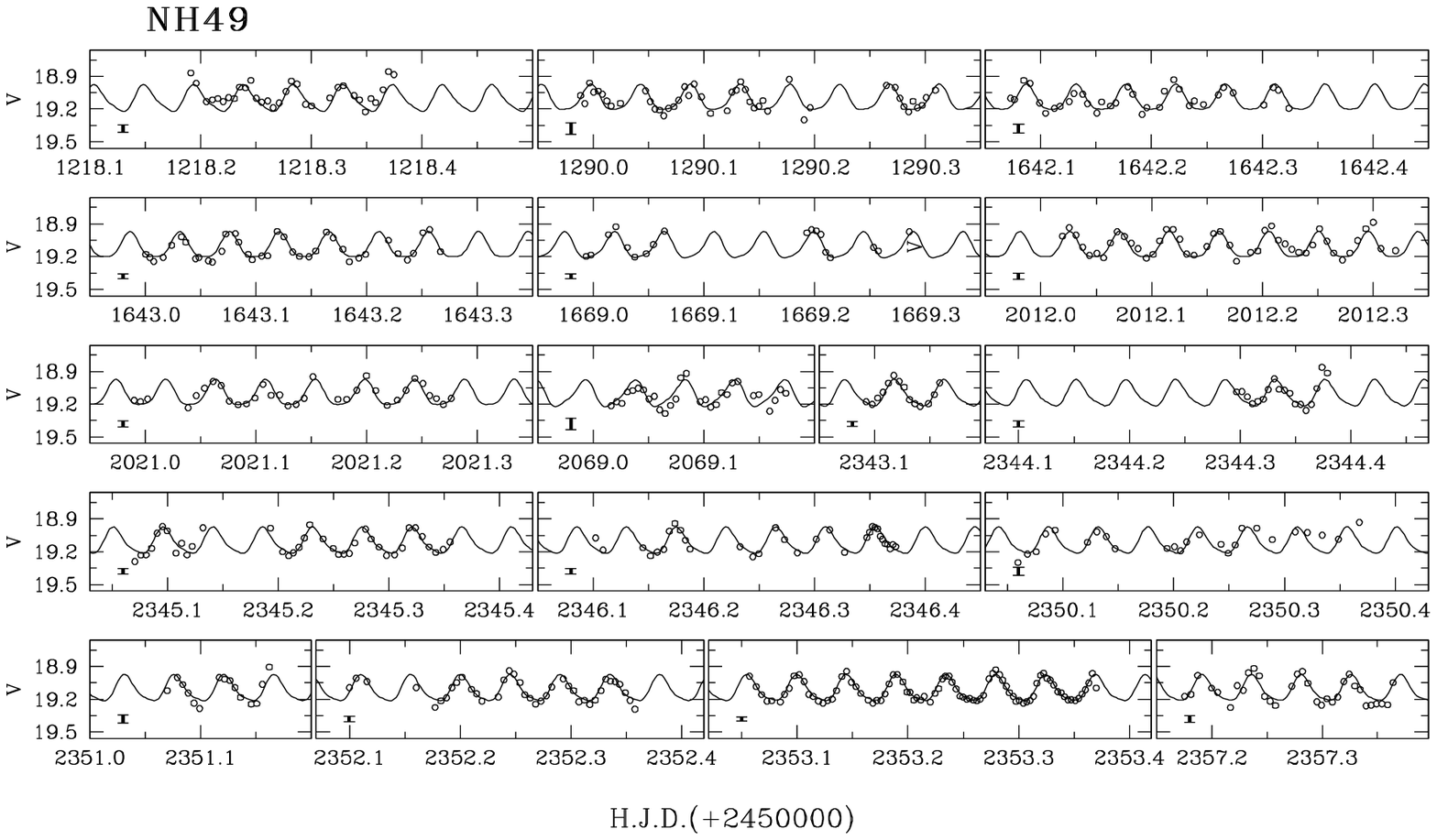}
\figcaption[Jeon.fig03i.ps]{ Observed data 
(circles) for nine SX Phoenicis stars.
Synthetic light curves (solid lines) obtained from the
multiple-frequency analysis (see Table 3) are superimposed on the data. 
We present the mean photomoetric errors of each observing days 
in the left lower corner of each panel.
\label{fig3}}

\clearpage

\plotone{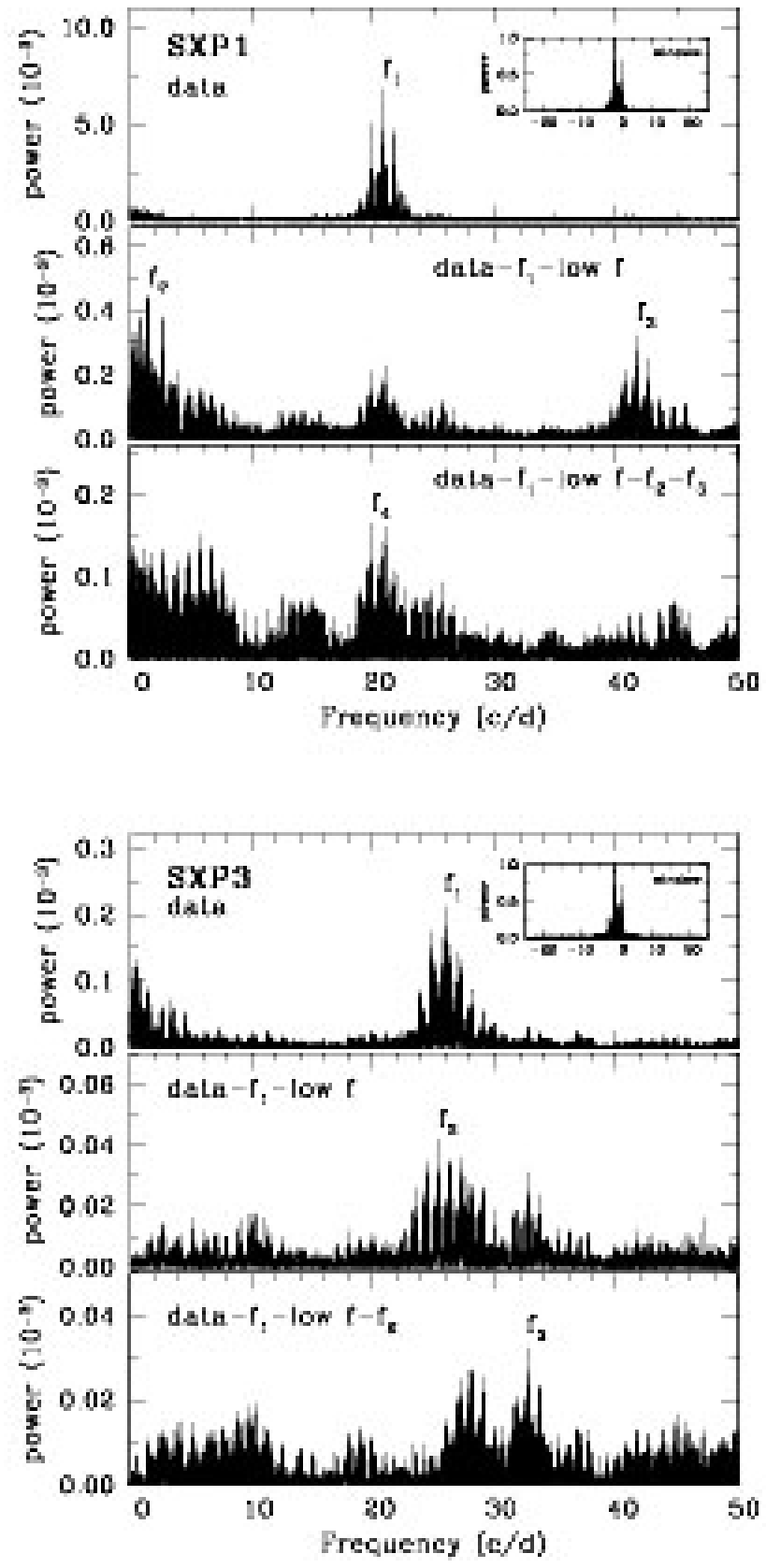}
\plotone{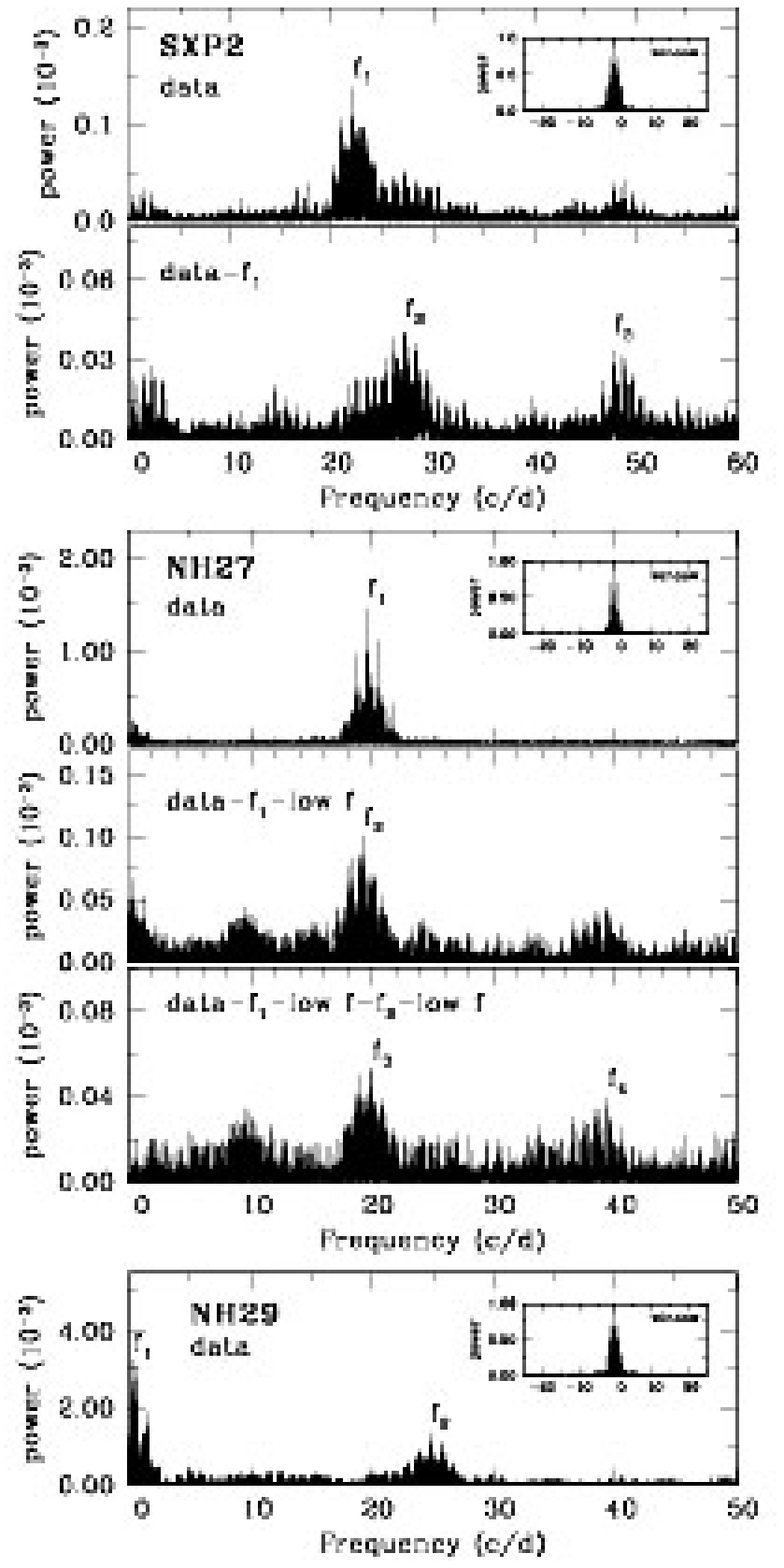}
\clearpage
\plotone{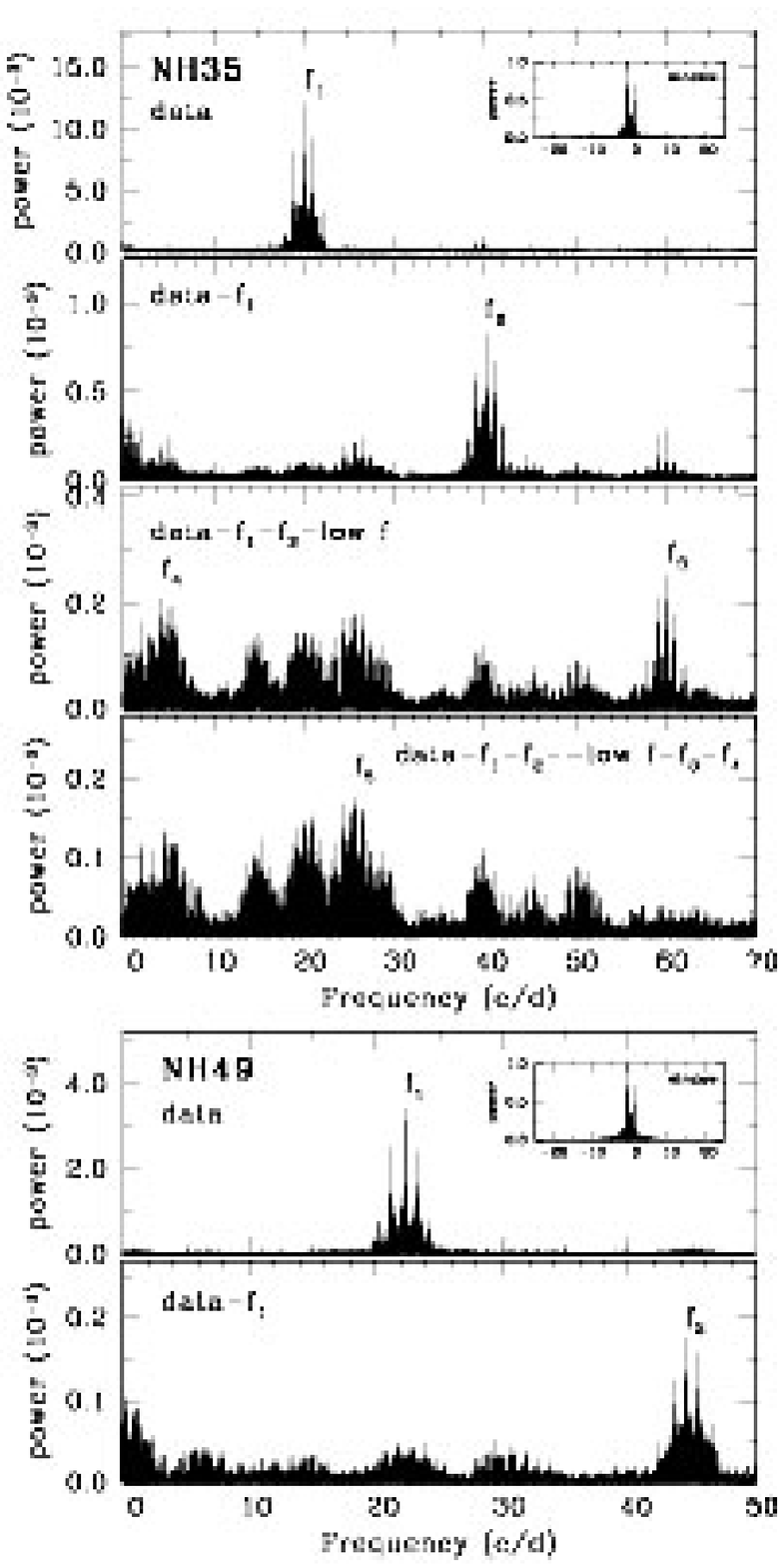}
\plotone{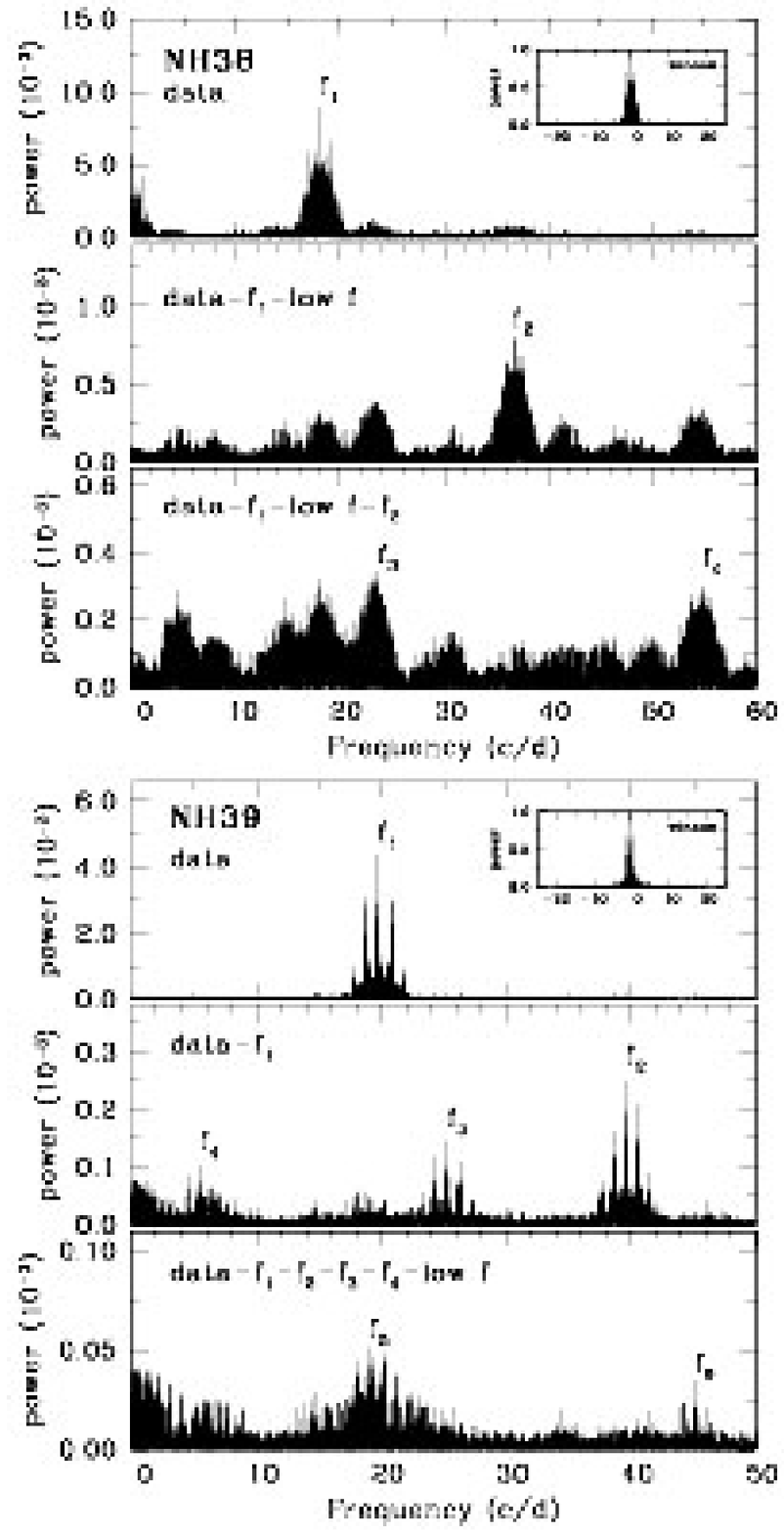}
\figcaption[Jeon.fig04bb.ps]{ Power spectra of
nine SX Phoenicis stars.
Window spectra are shown in a small box within each panel.
\label{fig4}}

\plotone{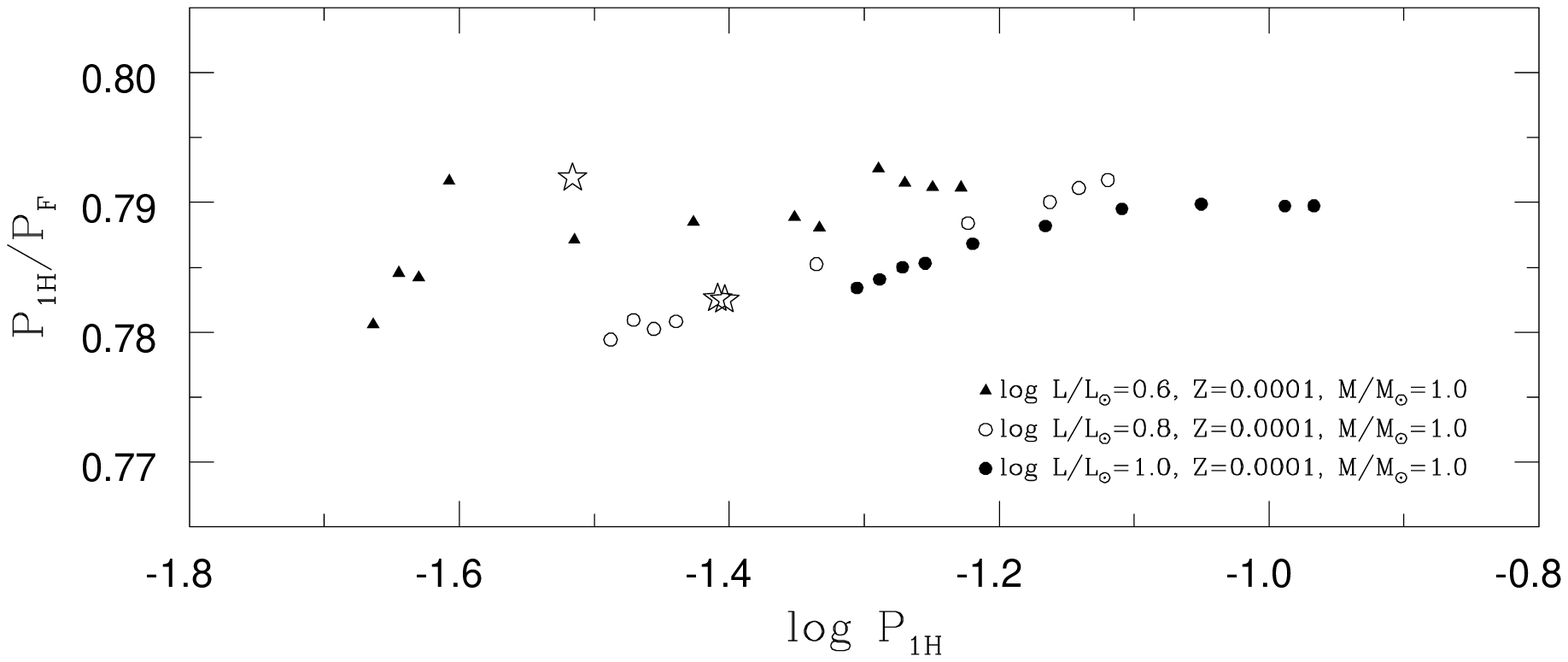}
\figcaption[Jeon.fig05.ps]{A diagram showing  the period of the first overtone mode versus
the period ratio of the the fundamental and first overtone modes.
The theoretical period ratios of the the fundamental and first overtone modes
are shown  for various $Log$ L/L$_\odot$ with Z=0.0001 and
M/M$_\odot$=1.0 given by Santolamazza et al. (2001).
Filled triangles, open circles and filled circles denote
$Log$ L/L$_\odot$ = 0.6, 0.8 and 1.0, respectively.
Star symbols are the double-radial mode SX Phoenicis stars in NGC 5466.
\label{fig5}}

\plotone{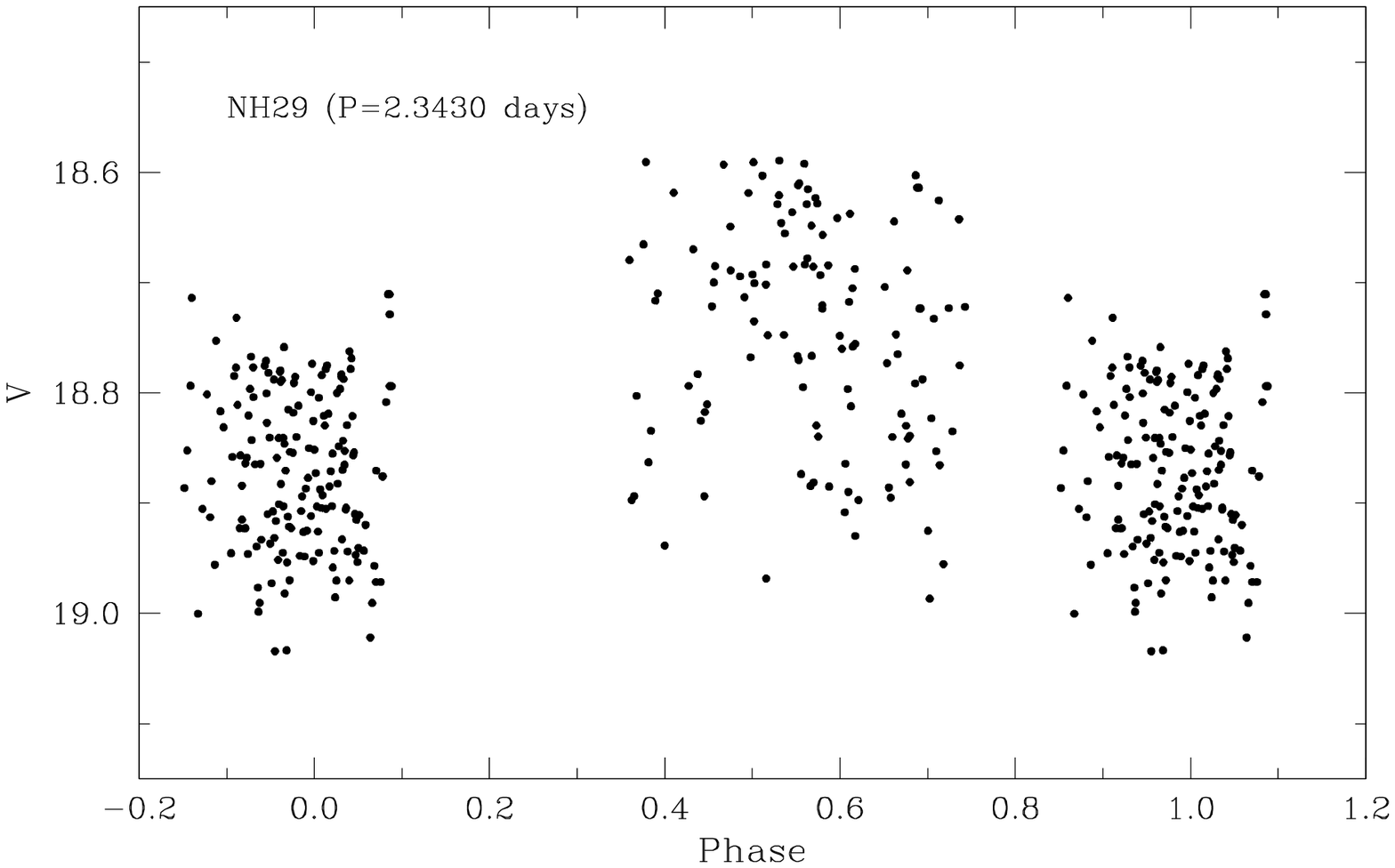}
\figcaption[Jeon.fig06.ps]{
A phase diagram for the long term variations of NH29.
Period and amplitude derived by multiple-frequency analyses
are 2.34302 days and 0.158 mag, respectively.
\label{fig6}}

\plotone{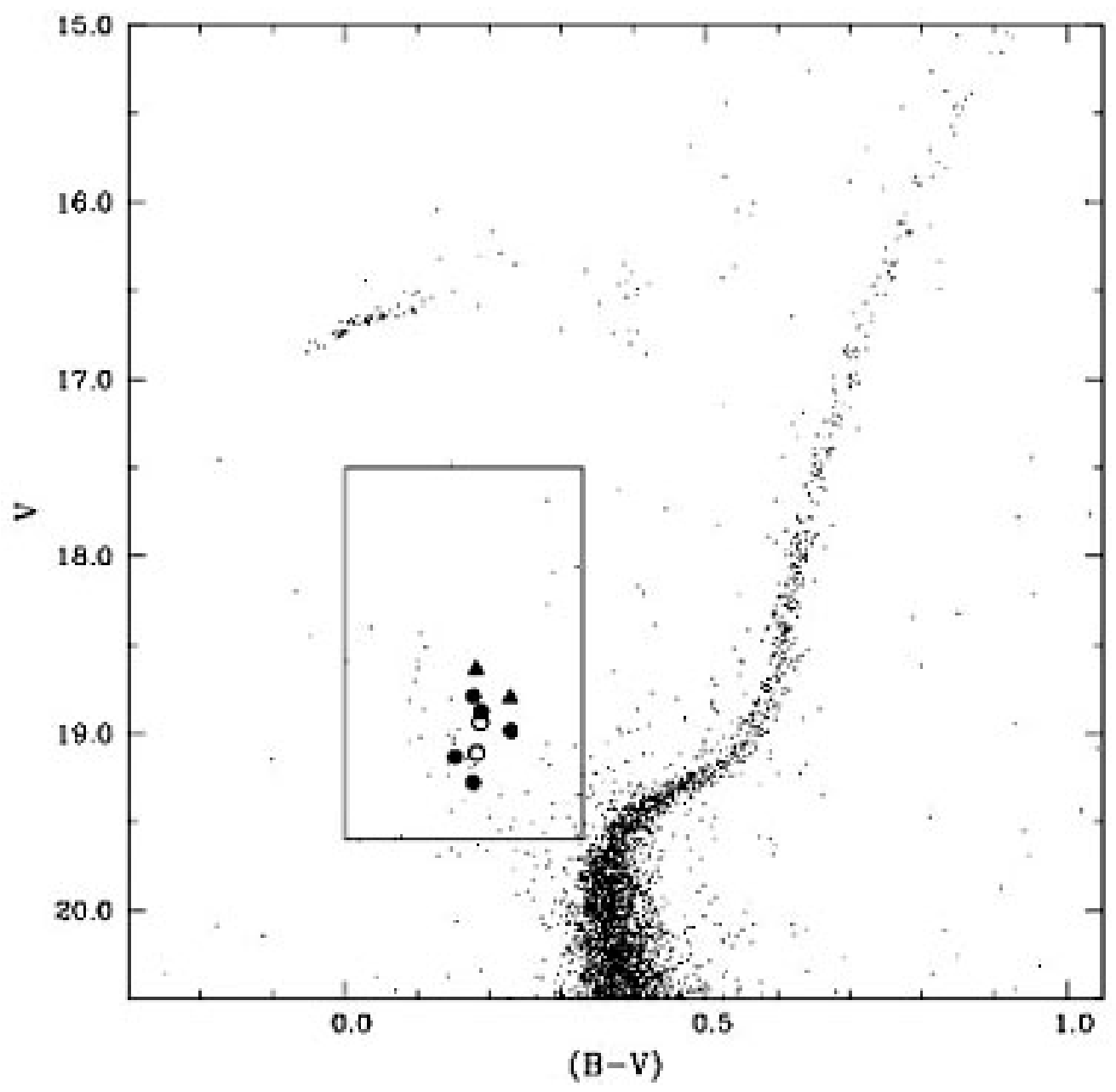}
\figcaption[Jeon.fig07.ps]{ Positions of  nine SX Phoenicis stars
in the color-magnitude diagram of NGC 5466. Note that all they are located in
the blue straggler region.
A filled triangle, filled circles and open circles denote the first overtone,
the double-radial mode of the fundamental and first overtone mode, and
the fundamental mode pulsators, respectively.
\label{fig7}}

\plotone{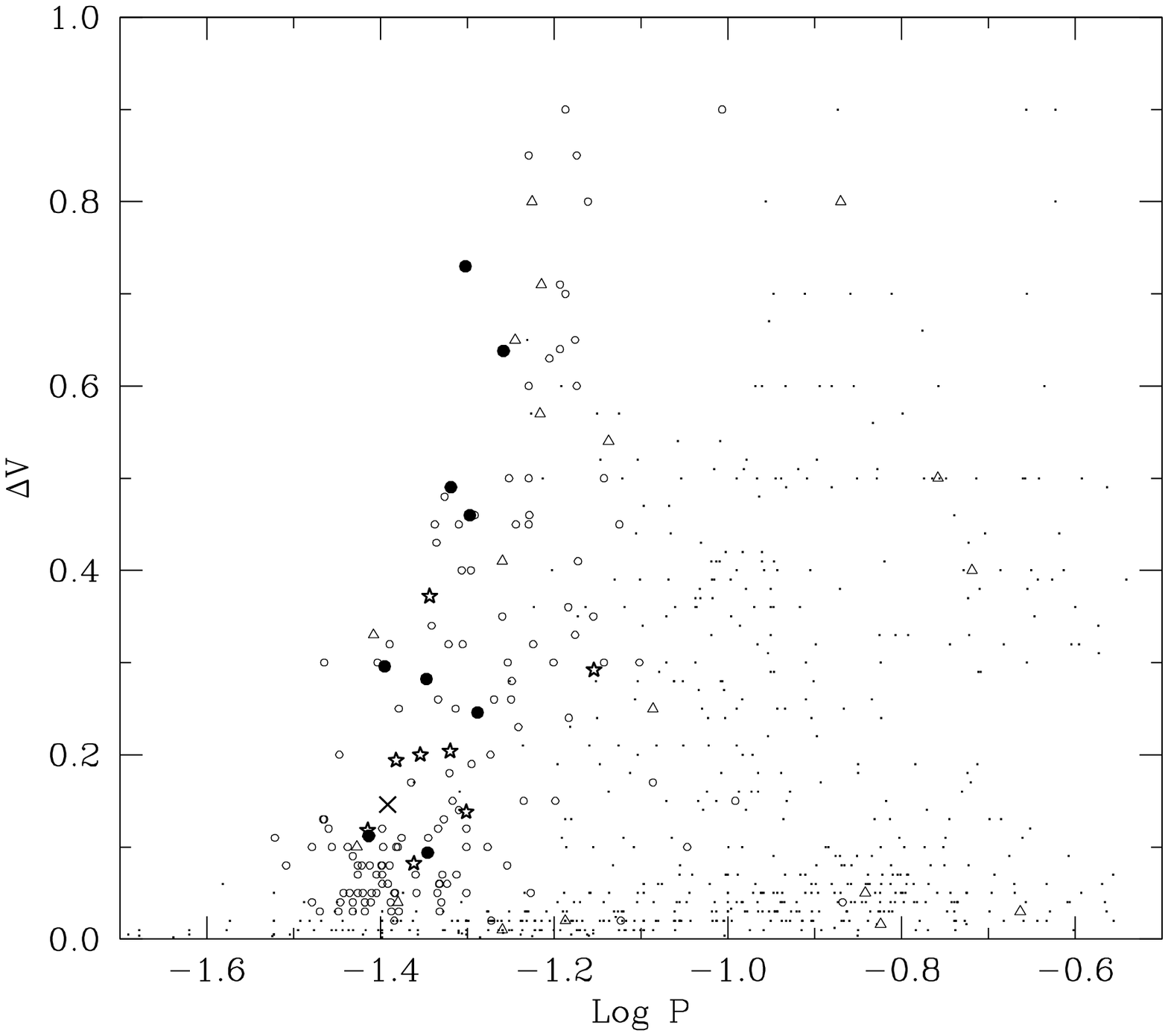}
\figcaption[Jeon.fig08.ps]{ $V$-amplitude versus  period diagram for the short period
pulsating stars.
Filled circles denote nine SX Phoenicis stars in NGC 5466.
Star symbols and a cross denote the SX Phoenicis stars in M53 and M15, respectively,
discovered in our previous search for variable blue
stragglers in globular clusters (Jeon et al. 2003; Jeon et al. 2001).
Triangles represent SX Phoenicis stars in the field,
 and open circles represent  SX Phoenicis stars in other globular clusters.
Dots denote $\delta$ Scuti stars.
\label{fig8}}

\plotone{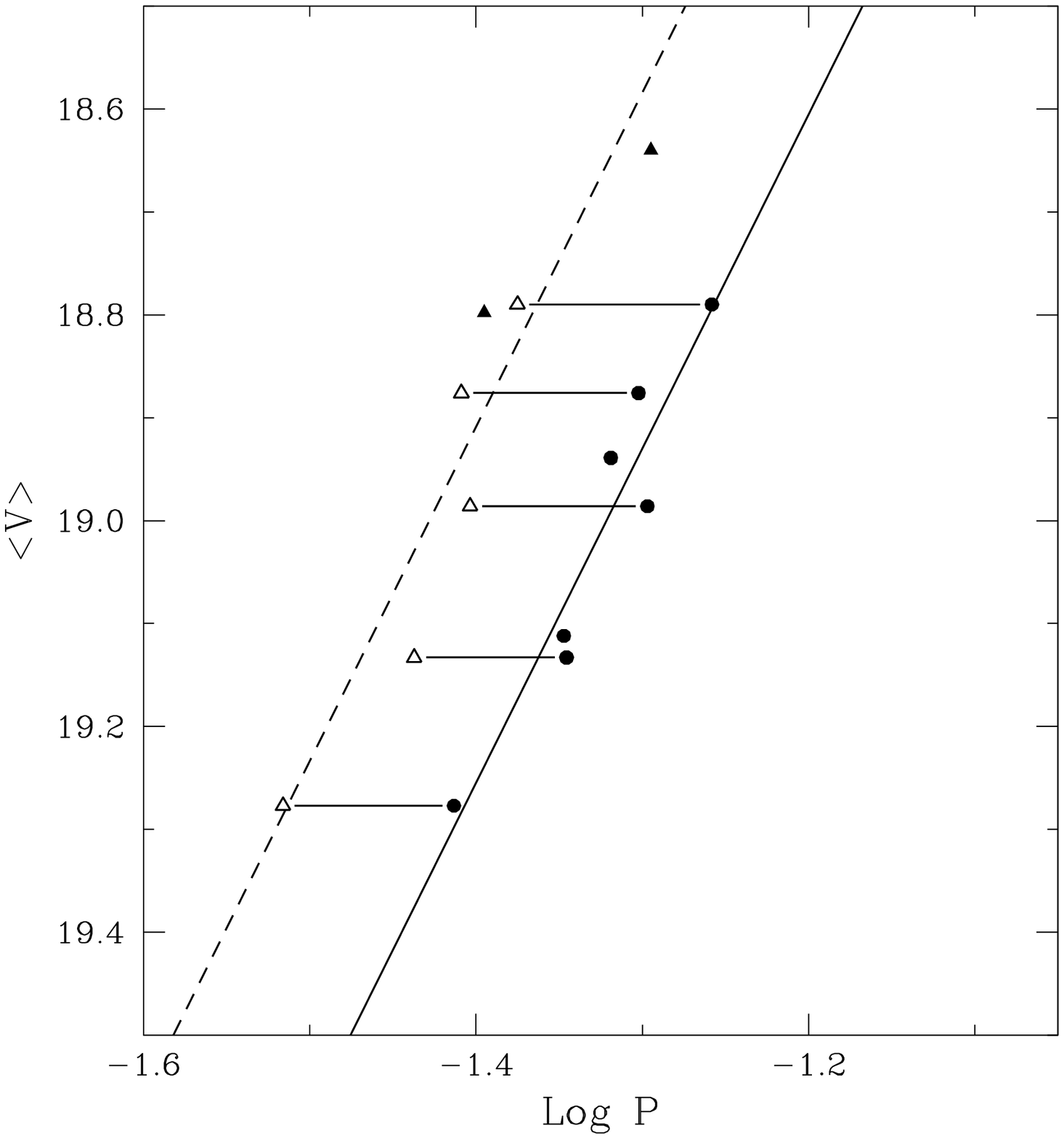}
\figcaption[Jeon.fig09.ps] {$ Log P$ versus mean magnitude $<$$V$$>$  diagram.
Filled circles and filled triangles denote the fundamental and first overtone modes
of SX Phoenicis stars in NGC 5466, respectively,
and open triangles show the first overtone mode elements for five
double-radial mode stars.
A solid line represents a P-L relation for the SX Phoenicis stars
with fundamental mode, and the line is shifted to a dashed line
corresponding to a P-L relation for the first overtone mode stars
by the ratio $P_{1H}/P_F = 0.783 $.
\label{fig9}}

\clearpage

\begin{deluxetable}{lrcrrrc}

\tablecaption{Observation Log. \label{tbl-1}} \tablewidth{0pt}
\tablecolumns{7}
\tablehead{
\colhead{Date} & \colhead{Start HJD} & \colhead{Duration} & \colhead{ $N_{obs}$  } &
\colhead{Seeing} & \colhead{Exposure Time} & \colhead{Remarks} \\
\colhead{(UT)} & \colhead{(2,450,000+)} & \colhead{(hours)} & \colhead{  } &
\colhead{(arcsec)} & \colhead{(seconds)} & \colhead{}
} \startdata
 1999~~2~~8  & 1218.191($V$) & 4.4 & 32($V$) & 1.5$\sim$1.9    & 300($V$)            &  \\
 1999~~3~27  & 1265.087($V$) & 6.2 & 56($V$) & 2.8$\sim$3.4    & 300($V$)            &  thin cloud, moon light\\
 1999~~4~~4  & 1273.031($V$) & 3.0 & 28($V$) & 2.5$\sim$3.0    & 300($V$)            &  moon light\\
 1999~~4~21  & 1289.980($V$) & 8.1 & 57($V$) & 1.8$\sim$2.2    & 300($V$)            &  \\
 2000~~4~~7  & 1642.038($V$) & 6.8 & 40($V$) & 2.4$\sim$3.0    & 180$\sim$210($V$) &                \\
 ~~~~~~~~~~  & 1642.044($B$) &     & 20($B$) &                 & 360($B$)            &  \\
 2000~~4~~8  & 1643.001($V$) & 6.7 & 42($V$) & 1.3$\sim$1.8    & 180($V$)            & \\
 ~~~~~~~~~~  & 1643.012($B$) &     & 18($B$) &                 & 360($B$)            &      \\
 2000~~4~10  & 1644.994($V$) & 7.6 & 51($V$) & 2.5$\sim$4.4    & 300($V$)            & thin cloud \\
 2000~~5~~4  & 1668.994($V$) & 7.0 & 18($V$) & 1.0$\sim$1.4    & 240($V$)            & many observing targets  \\
 ~~~~~~~~~~  & 1669.004($B$) &     &  4($B$) &                 & 360($B$)            & \\
 2001~~4~~7  & 2007.058($V$) & 6.6 & 72($V$) & 1.0$\sim$1.4    & 200($V$)            & full moon  \\
 2001~~4~12  & 2012.015($V$) & 7.3 & 49($V$) & 2.0$\sim$2.5    & 400($V$)            & moon light \\
 2001~~4~21  & 2020.990($V$) & 7.0 & 39($V$) & 2.0$\sim$2.5    & 400($V$)            & thin cloud \\
 2001~~6~~8  & 2069.016($V$) & 4.9 & 40($V$) & 1.5$\sim$2.3    & 300($V$)            & thin cloud \\
 2002~~3~~9  & 2343.092($V$) & 1.6 & 14($V$) & 1.8$\sim$2.2    & 300($V$)            & \\
 2002~~3~10  & 2344.067($V$) & 7.4 & 57($V$) & 2.5$\sim$4.5    & 300($V$)            & bad seeing \\
 2002~~3~11  & 2345.071($V$) & 7.0 & 51($V$) & 2.0$\sim$2.4    & 300$\sim$500($V$) & \\
 2002~~3~12  & 2346.103($V$) & 6.6 & 44($V$) & 0.9$\sim$3.0    & 120$\sim$300($V$) & many observing targets \\
 ~~~~~~~~~~  & 2346.239($B$) &     &  6($B$) &                 & 150$\sim$700($B$) & \\
 2002~~3~16  & 2350.045($V$) & 7.8 & 38($V$) & 2.5$\sim$3.0    & 300$\sim$600($V$) & many observing targets \\
 2002~~3~17  & 2351.071($V$) & 2.1 & 16($V$) & 2.0$\sim$2.3    & 300($V$)            & \\
 2002~~3~18  & 2352.100($V$) & 6.2 & 48($V$) & 1.8$\sim$3.2    & 250$\sim$400($V$) & \\
 2002~~3~19  & 2353.057($V$) & 7.5 & 86($V$) & 1.0$\sim$1.3    & 150$\sim$300($V$) & the best images \\
 2002~~3~22  & 2356.048($V$) & 7.3 & 30($V$) & 3.0$\sim$3.2    & 600($V$)            & thin cloud \\
 2002~~3~23  & 2357.176($V$) & 4.6 & 36($V$) & 2.3$\sim$3.0    & 300$\sim$400($V$) & \\
\enddata
\end{deluxetable}

\clearpage

\begin{deluxetable}{lcccc}

\tablecaption{Observational Parameters of the nine SX Phoenicis Stars. \label{tbl-2}} \tablewidth{0pt}
\tablecolumns{5}
\tablehead{
\colhead{Name} & \colhead{R.A.(J2000.0)} & \colhead{Decl.(J2000.0)} & \colhead{$<$$V$$>$ } &
\colhead{$<$$B$$>$$-$$<$$V$$>$}
} \startdata
Cl* NGC 5466 SXP 1 (New) &  14  5 39.25 &   28 31 18.5 &   18.939 &  0.188  \\
Cl* NGC 5466 SXP 2 (New) &  14  5 29.05 &   28 31 56.7 &   19.133 &  0.153  \\
Cl* NGC 5466 SXP 3 (New) &  14  5 28.12 &   28 34 49.1 &   19.277 &  0.178  \\
Cl* NGC 5466 NH 27 &  14  5 20.72 &   28 31 53.4 &   18.640 &  0.182  \\
Cl* NGC 5466 NH 29 &  14  5 23.61 &   28 31 37.9 &   18.798 &  0.229  \\
Cl* NGC 5466 NH 35 &  14  5 28.05 &   28 31  8.0 &   18.876 &  0.189  \\
Cl* NGC 5466 NH 38 &  14  5 28.10 &   28 32 37.7 &   18.790 &  0.178  \\
Cl* NGC 5466 NH 39 &  14  5 35.06 &   28 30 45.2 &   18.986 &  0.230  \\
Cl* NGC 5466 NH 49 &  14  5 25.86 &   28 31  2.7 &   19.112 &  0.183  \\

\enddata
\end{deluxetable}

\clearpage

\begin{deluxetable}{lcccrrcl}
\tablecaption{Pulsating Properties of the Nine SX Phoenicis Stars.
\label{tbl-3}} \tablewidth{0pt}
\tablecolumns{8} \tablehead{
\colhead{Name} & \colhead{Value} &
\colhead{Frequency\tablenotemark{{a,b}}} & \colhead{Amp.\tablenotemark{b}} &
\colhead{Phase\tablenotemark{b}}  &
\colhead{S/N\tablenotemark{c}} &  \colhead{Modes\tablenotemark{d}} &  \colhead{Remarks}
} \startdata
Cl* NGC 5466 SXP 1    &      $ f_1$& 20.8404 & 0.171   &  3.6823  & 34.0  & F\\
        &      $ f_2$&  1.5616 & 0.038   &  0.2057  &  8.7  & g ? & $\gamma$ Dor ? \\
        &      $ f_3$& 41.6809 & 0.036   &  3.3180  &  7.5  & $2f_1$\\
        &      $ f_4$& 19.9596 & 0.026   &  2.8318  &  5.3  & Nonradial$^d$ & \\ 
\cline{1-8}
Cl* NGC 5466 SXP 2       &  $ f_1$& 22.1600 & 0.023   &  0.9756  &  9.0  & F  & $f_1/f_2=0.810$\\
           &  $ f_2$& 27.3563 & 0.012   &--0.7726  &  4.9  & 1H$^e$~~?\\
           &  $ f_3$& 47.9119 & 0.012   &--1.5546  &  4.3  & $f_1+f_2$ ? &$f_1 + f_2 = 49.5$  \\
\cline{1-8}
Cl* NGC 5466 SXP 3      &   $ f_1$& 25.8994 & 0.031   &  3.6580  & 11.7  & F & $f_1/f_3=0.7919$\\
          &   $ f_2$& 25.4323 & 0.013   &  0.5014  &  5.5  & Nonradial \\
          &   $ f_3$& 32.7041 & 0.012   &  3.6235  &  4.6  &1H \\
\cline{1-8}
Cl* NGC 5466 NH 27       & $f_1$ & 19.7179 & 0.076    &  3.1794    & 24.7  & 1H or Nonradial & $f_1/f_3=0.980$\\
           & $f_2$ & 19.4181 & 0.020    &  1.1587    &  6.5  & Nonradial \\
           & $f_3$ & 20.1126 & 0.015    &  4.3056    &  4.7  & Nonradial \\
           & $f_4$ & 39.4356 & 0.012    &  0.3457    &  4.0  & 2$f_2$    \\
\cline{1-8}
Cl* NGC 5466 NH 29    &      $ f_1$&  0.4268 & 0.079   &  3.9715  & 18.2  & Eclipse ?\\
        &      $ f_2$& 24.8362 & 0.069   &  0.7290  & 11.3  & 1H\\
\cline{1-8}
Cl* NGC 5466 NH 35        &  $ f_1$& 20.0583 & 0.221   &  2.7076  & 41.1  & F  &$f_1/f_5=0.7826$  \\
        &      $ f_2$& 40.1167 & 0.056   &  1.6236  & 10.7  & $2f_1$\\
        &      $ f_3$& 60.1750 & 0.032   &  2.3160  &  5.9  & $3f_1$\\
        &      $ f_4$&  4.3640 & 0.029   &  3.8185  &  5.4  & $f_5-f_1$\\
        &      $ f_5$& 25.6302 & 0.027   &  0.8393  &  5.0  & 1H\\
\cline{1-8}
Cl* NGC 5466 NH 38    &      $ f_1$& 18.1207 & 0.189   &  4.4947  & 25.9  & F & $f_1/f_3=0.764$\\
        &      $ f_2$& 36.6623 & 0.055   &--0.3926  &  7.7  & $2f_1$\\
        &      $ f_3$& 23.7158 & 0.037   &  4.0499  &  5.1  & 1H$^e$~~?   \\
        &      $ f_4$& 54.7930 & 0.038   &  1.5403  &  4.7  & 3$f_1$ \\
\cline{1-8}
Cl* NGC 5466 NH 39       &   $ f_1$& 19.8124 & 0.128   &  1.1796  & 46.6  & F & $f_1/f_3=0.7825$  \\
        &      $ f_2$& 39.6248 & 0.032   & --0.2177 & 11.2  & $2f_1$\\
        &      $ f_3$& 25.3187 & 0.024   &  4.1147  &  8.2  & 1H\\
        &      $ f_4$&  5.5092 & 0.019   &  0.4269  &  7.0  & $f_3-f_1$\\
        &      $ f_5$& 19.0786 & 0.015   &  0.6885  &  5.1  & Nonradial & $f_5/f_1=0.963$  \\
        &      $ f_6$& 45.1339 & 0.012   & --1.1017 &  4.1  & $f_1+f_3$\\
\cline{1-8}
Cl* NGC 5466 NH 49     &  $ f_1$& 22.2415 & 0.115   &  0.1169  & 37.6  & F \\
         &      $ f_2$& 44.4803 & 0.026   & --0.1681  &  8.5  & $2f_1$\\

\enddata

\tablenotetext{a}{In cycles per day. }
\tablenotetext{b}{$V = Const + \Sigma_j A_j \cos \{2 \pi
          f_j (t - t_0) + \phi_j\},~~ t_0 =$ HJD 2,450,000.00. }
\tablenotetext{c}{S/N = $\rm{\{(power~~for~~each~~frequency)~~/~~(average~~power~~from~~0~~\sim~~70~~cycle~~day^{-1}}~~
after~~prewhitening~~all~~frequencies)\}^{1/2}$}
\tablenotetext{d}{F \& 1H : fundamental and first overtone radial mode, respectively; g : gravitational nonradial mode;
                  $2f_1$, $3f_1$ \& $f_3-f_1$, etc. : combination frequencies}
\tablenotetext{e}{Probably affected by 1 cycle day$^{-1}$ alias. }

\end{deluxetable}
\clearpage

\end{document}